\documentstyle[11pt,aaspp4,flushrt,figwithcaption]{article}

\newcommand{\Ham}{{\cal H}}	
\newcommand{\Lag}{{\cal L}}	
\newcommand{\Nbody}{$N$-body}
\newcommand{\PM}{PM}
\newcommand{\PP}{PP}

\newcommand{\PMM}{PM$^2$}
\newcommand{\PPPM}{P$^3$M}
\newcommand{\APPPM}{AP$^3$M}
\newcommand{\eg}{{\em e.g.}}

\newcommand{\adda}{\frac{\ddot{a}}{a}} 
\newcommand{\half}{\frac{1}{2}}		
\newcommand{\Ncells}{n_c}
\newcommand{\Neff}{N_{\rm eff}}
\newcommand{\thrdL}{\frac{\Lambda}{3}}	
\newcommand{\rhob}{\rho_b}
\newcommand{\rhobar}{\bar\rho}

\newcommand{\norm}[1]{|#1|}	
\newcommand{\vsq}[1]{\norm{#1}^2}  
\renewcommand{\vec}[1]{{\bf #1}}	

\newcommand{\dtau}{\delta\tau}
\newcommand{\hdtau}{\dtau/2}
\newcommand{\vecF}{\vec{F}}
\newcommand{\vecJ}{\vec{J}}
\newcommand{\vecP}{\vec{P}}
\newcommand{\vecf}{\vec{f}}
\newcommand{\vecp}{\vec{p}}
\newcommand{\vecr}{\vec{r}}
\newcommand{\vecrd}{\dot{\vecr}}

\newcommand{\vecx}{\vec{x}}
\newcommand{\vecxd}{\dot{\vecx}}
\newcommand{\vecxdd}{\ddot{\vecx}}

\received{1996 June 6} 
\revised{1996 October 29}  
\accepted{1996 November 19} 
\journalid{480}{1997 May 1}
\paperid{34692}
\slugcomment{ApJ 480, in press}

\lefthead{Gelato, Chernoff, and Wasserman}
\righthead{Adaptive Hierarchical Particle-Mesh}

\begin{document}

\title{An Adaptive Hierarchical Particle-Mesh Code With Isolated
Boundary Conditions}
\author{Sergio Gelato}
\affil{SISSA, via Beirut 2--4, I--34013 Trieste, Italy}
\author{David F. Chernoff and Ira Wasserman}
\affil{Center for Radiophysics and Space Research, 
Cornell University, Ithaca NY 14853--6801, USA}

\begin{abstract}
This article describes a new, fully adaptive Particle-Multiple-Mesh
(\PMM{}) numerical simulation code developed primarily for
cosmological applications.
The code integrates the equations of motion of a set of particles
subject to their mutual gravitational interaction and to an optional,
arbitrary external field.
The interactions between particles are computed using a hierarchy of
nested grids constructed anew at each integration step to enhance the
spatial resolution in high-density regions of interest.
As the code is aimed at simulations of relatively small volumes of
space (not much larger than a single group of galaxies) with
independent control over the external tidal fields, significant effort
has gone into supporting isolated boundary conditions at the top grid
level.
This makes our method also applicable to non-cosmological problems,
at the cost of some complications which we discuss.
We point out the implications of some differences between our approach
and those of other authors of similar codes, in particular with
respect to the handling of the interface between regions of different
spatial resolution.
We present a selection of tests performed to verify the correctness
and performance of our implementation.
The conclusion suggests possible further improvements in the areas of
independent time steps and particle softening lengths.
\end{abstract}


\section{Introduction}

Ideally, a good cosmological \Nbody{} simulation should resolve in
some detail the internal structure of individual galaxies (on mass
scales as low as $10^6$ to $10^7\,M_{\sun}$ and on length scales of
a few kiloparsecs or less) while at the same time representing the
growth of density perturbations on the scale of clusters and even
superclusters of galaxies.
Galaxy morphologies are observed to correlate with their membership in
clusters.
The Virgo cluster is estimated to be an appreciable, if not dominant,
source of tides on the Local Group 15 to 20~Mpc away.
And when studying the nonlinear growth of galaxy-sized perturbations,
one should not discount the possible coupling to modes with
wavelengths as large as 50~Mpc (\cite{GB94}).

Achieving the required dynamic range is a
difficult task given the limitations of present-day computers.
Generally, numerical codes have been able to obtain a large dynamic
range either in mass or in length, but not in both.
Particle-particle (\PP) methods (\cite{Aa85}) and tree codes (Appel
\nocite{Ap81}1981, \nocite{Ap85}1985; \cite{BH86}; \cite{JP89}) have
essentially unlimited length resolution, bounded only by the need to
soften two-body encounters when modeling a collisionless physical
system, but their use for simulations with more than a few tens,
respectively hundreds, of thousands of particles is currently
restricted to special-purpose hardware (such as GRAPE chips) or massively
parallel computers (\cite{Du96}).
Particle-mesh (\PM) methods (\cite{HE81}, hereafter HE;
\cite{EDFW85}), by contrast,
can integrate the orbits of millions of interacting particles
even on a more modest single-processor mainframe or high-end workstation.
The forces are computed by solving Poisson's equation on a Cartesian
grid. 
Barring complications such as a memory latency that increases with
problem size,
the computational cost per time step is linear in the number of
particles and of order $N \log N$ in the number of grid cells (which
is typically of the same order as the number of particles).
The dynamic range in length, however, is limited by the maximum size
of the grid, typically 256 or 512 cells on a side in three dimensions
on a current single-processor computer.
Parallel processing can naturally push this limit towards higher
values.

One can circumvent this obstacle by decomposing the inter-particle
forces into a long-range part, calculated by the \PM{} technique, and
a small-range part which can be evaluated in a number of different
ways.
For some time, a popular approach has been to compute the short-range
forces by direct summation over all sufficiently close pairs of
particles.
This technique, known as the particle-particle, particle-mesh (\PPPM)
method (HE \S 8), was originally applied to plasma simulations
where the electrostatic repulsion between charges of the same sign
makes it difficult for large density contrasts to develop.
Its use with gravitating systems suffers from the tendency for ever
more particles to condense into small volumes, causing the cost of the
\PP{} summation to become prohibitive as the system evolves.
Increasing the total number of particles and resolving smaller scales
(on which density perturbations become nonlinear at earlier times, at
least in the presently favored ``bottom-up'' scenario) both exacerbate
the problem.
We are unaware of any \PPPM{} simulations using more than about $5\times
10^5$ particles.
One may be able to raise this limit by using a tree method to compute
the short-range forces, as in Xu's \nocite{Xu95}(1995) TPM, but in
principle the difficulty remains.

Most other attempts to enhance the spatial resolution of \PM{} codes
rely on the introduction of local grid refinements.
This idea lies at the root of Couchman's (1991)\nocite{Co91} adaptive
\PPPM{} (\APPPM) (which however still uses direct summation where the
number of particles is small enough), of numerous hierarchical \PM{} codes
(\cite{C.86}; \cite{Vi89}; \cite{BFW91}; \cite{ANC94}, hereafter ANC;
\cite{GCW94}; \cite{JDC94}, hereafter JDC; \cite{Sp96})
and of various related approaches (\cite{JW86}; \cite{SS95a}).

As pointed out by various authors, and most recently by Suisalu and
Saar (1995b)\nocite{SS95b}, the choice between \PM{}, \PPPM{}, and
tree codes is not only a question of the spatial and mass resolution
that can be achieved, but also of the behavior of each method with
respect to two-body gravitational collisions.
These are to be avoided when the physical system being modeled is
essentially collisionless, as is the case for example for cosmological
dark matter.
\PM{} codes usually prove to be less collisional, although it should
be emphasized that this is at least in part a consequence of their
spatial resolution being weaker than their mass resolution, and need
not carry over to hierarchical \PM{} methods unless precautions are taken
to ensure that the mass granularity remains adequate at all times.

The purpose of this article is to document in some detail our
implementation of a dynamically adaptive multiple-mesh code and
the tests we performed to validate it.
So far, only a brief report of an earlier stage of development
(\cite{GCW94}) and a description of the methodology but not of the
tests (\cite{Ge95}) have appeared.
Our code, like that of JDC,
is able to track the formation and subsequent motion of density
concentrations by dynamically adjusting the number, nesting depth and
location of the subgrids.
In our case, the entire grid structure is chosen afresh on every step.
Since our first intended application was a cosmological problem---the
formation of the Local Group---that requires the freedom to specify
external tidal fields other than would be implied by periodic image
charges, our code presents the relatively uncommon combination of an
expanding system of coordinates with isolating (more properly
non-periodic) boundary conditions.
This introduces a number of complications not usually discussed in the
literature on \PM{} codes, and of which we shall give an account here.
(Our method is also applicable to non-cosmological problems, for which
isolating boundary conditions are an asset and
the aforementioned complications do not arise.)
A future article (\cite{GCW96b}) will detail the results of our
simulations of group formation.

The general outline of this paper is as follows.
Section~\ref{s:method} describes in detail the method used.
The tests and their results are presented in section~\ref{s:tests}.
We conclude, in section~\ref{s:conclusions}, with our assessment of
the strengths and limitations of this code.

\section{Description of the method}
\label{s:method}

\subsection{Equations of motion}

Our code integrates the equations of motion for a set of gravitating
particles.
These equations can be derived from the Lagrangian
\begin{equation}
\label{q:Lag-r}
\Lag = \half \sum_{i} m_{i} \vsq{\vecrd_{i}} +
\half \sum_{i} \sum_{j\ne i} m_{i} m_{j} V(\vecr_i, \vecr_j; t)
+ \half \thrdL \sum_{i} m_{i} \vsq{\vecr_{i}}
- \sum_{i} m_{i} \Phi_x(\vecr_i; t).
\end{equation}
Here the indices $i$ and $j$ span the set of particles in the system,
$m_i$ is the mass of particle $i$, $\vecr_i(t)$ its position at time
$t$,
and $V(\vecr,\vecr';t)$ describes the law of pairwise interaction
between particles.
Ideally $V(\vecr,\vecr';t)=1/\norm{\vecr-\vecr'}$ (we choose units such
that $G=1$) but the numerical method forces us to use an approximation
which, due to adaptive grid refinement, can depend explicitly on time
and on the individual coordinates $\vecr$ and $\vecr'$, not just on
their difference.

We could have included the term proportional to $\Lambda$ (where
$\Lambda$ is an arbitrary constant) within the external potential
$\Phi_x$, but for cosmological applications it is useful to show it
explicitly. 

\subsection{Expanding coordinates}

It is often convenient to apply a time-dependent rescaling of the
coordinate system, with new coordinates $\vecx$ defined by
$\vecr=a(t)\vecx$.
An equivalent Lagrangian is then
\begin{equation}
\label{q:Lag-x}
\Lag = \half a^2 \sum_i m_i \vsq{\vecxd_i} 
+ \half a^2 \left(\thrdL - \frac{\ddot{a}}{a} \right)
\sum_i m_i \vsq{\vecx_i}
+ \half \sum_i \sum_{j\ne i} m_i m_j V(\vecx_i, \vecx_j; t)
- \sum_i m_i \Phi_x (\vecx_i; t).
\end{equation}
For notational convenience, we define
\begin{eqnarray}
H & \equiv & \frac{\dot{a}}{a} \\
\rhobar & \equiv & \frac{3}{4\pi} \left( \thrdL - \adda
\right) .
\end{eqnarray}
Note that in principle $a(t)$ is arbitrary, reflecting our freedom to
choose the coordinate system.
For cosmological applications one normally chooses $a(t)$ to be the solution of
Friedman's equation for some cosmological model, in which case
$\rhobar(t)$ is the background density and $H(t)$ the
Hubble constant for that model.
Later in this paper we shall impose the additional restriction
$\rhobar a^3 = {\rm constant}$, which holds for cosmological models in
the matter-dominated era.

The second term in the Lagrangian~(\ref{q:Lag-x}) can be regarded as
involving an effective potential
\begin{equation}
\Phi_a (\vecx; t) = - \half a^2 \vsq{\vecx} \left(\thrdL - \adda \right)
= - \frac{2\pi}{3} \rhobar a^2 \vsq{\vecx},
\end{equation}
which satisfies
$\nabla^2 \Phi_a = - 4\pi \rhobar a^2$.
Similarly, in the continuum limit and for a Coulomb interaction
($V(\vecx,\vecx';t)=a^{-1}(t)\norm{\vecx-\vecx'}^{-1}$), 
the third term satisfies 
$\nabla_2^2 \sum_j m_j V(\vecx_i,\vecx_j;t) = -4\pi \rho(\vecx_i) a^2$, 
where 
the proper density $\rho$ is given by
$\rho(\vecx) a^3=\sum_j m_j \delta(\vecx-\vecx_j)$
and
$\nabla_1$ (respectively $\nabla_2$) denotes differentiation with
respect to the first (respectively the second) of the 
two position vectors on which $V$ depends.

The usual derivation of the equations of motion from the
Lagrangian~(\ref{q:Lag-x}) yields:
\begin{equation}
\label{q:motion}
\vecxdd_i + 2 H \vecxd_i = \frac{4\pi}{3}\rhobar \vecx_i
+ a^{-2} \sum_{j\ne i} m_j \nabla_1 V(\vecx_i, \vecx_j; t)
- a^{-2} \nabla \Phi_x(\vecx_i; t).
\end{equation}
For later convenience, we define
\begin{equation}
\vecf_i = \frac{4\pi}{3} \rhobar a^3 \vecx_i
+ a\sum_{j\ne i} m_j \nabla_1 V(\vecx_i,\vecx_j;t) 
- a\nabla\Phi_x (\vecx_i; t).
\end{equation}
This allows the equation of motion~(\ref{q:motion}) to be written more
compactly as
\begin{equation}
\vecxdd_i + 2 H \vecxd_i = a^{-3} \vecf_i .
\end{equation}
It is also useful to recast the equation in terms of a generalized
time coordinate $\tau(t)$:
\begin{equation}
\frac{d^2\vecx_i}{d\tau^2} + 2 A(\tau) \frac{d\vecx_i}{d\tau} =
B(\tau) \vecf_i
\end{equation}
with
\begin{eqnarray}
\label{q:A}
A(\tau) & \equiv & \half \left(\frac{dt}{d\tau}\right)^2
\left(\frac{d^2\tau}{dt^2}+2H\frac{d\tau}{dt}\right) \\
\label{q:B}
B(\tau) & \equiv &
\frac{1}{a^3} \left(\frac{dt}{d\tau}\right)^2 .
\end{eqnarray}
A common choice is the so-called conformal time, $d\tau=a^{-1}dt$;
another (\cite{EDFW85}) is $\tau=a^{\alpha}$ for some constant
$\alpha$.
Naturally, any sufficiently differentiable monotonic function of~$t$
is acceptable, and one is 
free to construct such a function {\it ad hoc} for each individual
simulation. 
We avail ourselves of this freedom.

\subsection{The top grid}

We lay down a rectangular grid of $N_x$ by $N_y$ by $N_z$ nodes with
uniform spacings $h_x$, $h_y$ and~$h_z$.
The values of $N_x$, $N_y$ and $N_z$ must be acceptable to the Fast
Fourier Transform (FFT) routines used.
(All our tests were done with $N_x=N_y=N_z$ a power of 2, and with
$h_x=h_y=h_z$, as this is the most common situation in practice and
the only one we needed for our applications.)
At every step, a density is assigned to each grid point using the
cloud-in-cell (CIC) algorithm.
FFTs are then used to solve for the potential on the same grid, and
an acceleration is obtained for each particle by differentiating  the
grid potential, then using CIC interpolation.
We use a two-point finite-difference formula to compute the
derivatives of the potential at grid nodes.
This choice means that truncation errors in the force law scale with
the square of the grid spacing (HE, \S 5-4).

The discrete Green's function we use is obtained by sampling
$1/\norm{\vecx-\vecx'}$ at grid points (and Fourier transforming the
result).
This differs from the more common approach in cosmological \PM{} codes
(\cite{EDFW85,Vi89}) of basing the Green's function on the seven-point
finite difference approximation to the Laplacian; our choice has the
merit of guaranteeing the truncation of the interaction law at large
separations, as required for a correct implementation of isolated boundary
conditions (\cite{EB79}).
Unlike in the \PPPM{} method, where the grid-based part of the force
must possess a good degree of translational invariance
to avoid introducing terms in the direct particle-particle
contribution that depend explicitly on the position of each pair
relative to the grid,
we are not compelled to soften the interaction at small scales.
To do so would cost us precious spatial resolution; already with our
choice of Green's function the effective force turns out to be
softened at separations $\lesssim 2.5$ grid cells.
A consequence of our decision not to soften the force law any further
is that the $Q$-minimization procedure of HE \S 8-3-3, which requires
the reference force to have no harmonics on scales smaller than the
grid spacing, is not appropriate for us.
The choice of the shape of the interaction law at small separations
and of the charge assignment and force interpolation scheme is a
matter of compromise between accuracy and computational cost.
Our choices could undoubtedly be improved upon, but we preferred to
concentrate our efforts on the more innovative aspects of our method.

\subsection{Boundary conditions}
\label{s:bc}

Isolated boundary conditions are implemented by conceptually doubling
the size of the grid in each dimension, and padding the density array
with an appropriate constant value (normally zero) outside the
principal octant, which contains the particles (HE, \S 6-5-4).
If the Green's function is truncated so that the interaction falls to
zero at separations larger than the system (before doubling), this
completely suppresses any interaction between the system and its periodic
images.
The computed potential, however, will only match that of an isolated
system within the principal octant.
Since we need to apply a gradient operator to the potential in order
to compute the forces, there is a layer (one cell thick for our
approximation to the gradient) in which we would not be able to
compute them. 
We therefore allow no particles in this outer layer.
(We are free to increase the thickness of this layer.
It has been convenient to do so in some of the tests discussed below,
to achieve an integral ratio of cell counts in comparisons between
different grid resolutions.)
We do not use James' \nocite{Ja77} (1977) method to impose isolated
boundary conditions without doubling the grid by calculating
appropriate screening charges on the surface, since strictly speaking
the procedure is only justified when a finite-difference
approximation to the Laplacian is used in solving Poisson's equation.
The FFT technique does not satisfy this condition, and would require
screening charges throughout the volume of the box.

In cosmological applications, the space surrounding the system is not
to be thought of as empty, but rather as containing a background of
uniform density $\rhob$.
(Departures from uniformity can be represented through an appropriate
choice for the tidal potential $\Phi_x$.)
We must therefore add this background, convolved with
the same (CIC) charge assignment function that is used for the particles.
Its contribution to the density at any node on the boundary of the
region where particles are allowed is then proportional to the number
of cells adjacent to this node that lie outside the particle region:
$\rhob/2$ for nodes on a face, $3\rhob/4$ on an edge, $7\rhob/8$ at a
vertex.

We implement the $(4\pi/3)\rhobar\vecx$ term of the equation of
motion (equation~\ref{q:motion}) by subtracting $\rhobar$ from the
density at all points before solving for the potential.
If $\rhobar$ coincides with the background cosmological density, the
source function for the potential outside the principal octant is
exactly zero. 
This is the most common case, and the one that presents
the fewest conceptual and practical difficulties.
With periodic boundary conditions it would also be the only possible
case since the mean density outside the simulation box must always
equal that inside it.
With isolated conditions a mismatch is permissible, and could be used for
example in applying an expanding or contracting grid to follow the
evolution of an isolated system without a cosmological density
background.
The varying $a(t)$ would then be adjusted to match the expansion or
contraction of the simulated system, providing better resolution at
lower cost during the collapse phase.

A very significant difference between periodic and isolated boundary
conditions is that the latter allow the exchange of mass, momentum,
energy, angular momentum between the particles and the exterior.
With periodic boundary conditions, there is effectively no exterior.
Particle flows are a significant concern in cosmological applications,
where matter may both leave and enter the computational box during the
simulation.
In a typical cosmological model, the mass variance is of order unity
in a sphere of radius $r_8=8 h^{-1}$~Mpc.
One expects particles at the edge of a sphere of radius $r_8$ to be
displaced with respect to the center of mass of the sphere by about
$r_8/3$, a significant length when compared to the
size of a group or cluster of galaxies.
A simulation of characteristic size $r_8$ may reasonably be expected
to exhibit a twofold increase or decrease in the total mass within
the box during a run.
On smaller scales the variance is even larger, at least for the
currently fashionable ``bottom-up'' scenarios of structure formation.

Exiting particles are easily handled by removing them from the
simulation, but incoming particles have to be injected according to
some prescription that fits the physical problem at hand.
In the cosmological case, as long as the density perturbations grow
linearly on the scale of the box, a reasonable prescription can be
based on extrapolations from the linear theory.
At late times, in the strongly non-linear regime, simply extrapolating
from the linear solution causes large amounts of material
to be injected which
in reality would collapse into bound objects outside the box and
remain outside. In other words, the Zel'dovich approximation is clearly
inappropriate beyond the time at which caustics form. It would lead to
a gross overestimate of the total mass flow into the box.
An adequate model of the inflow in the nonlinear regime therefore
requires an actual simulation of the mass flows in a larger region.
This has become normal practice in simulations with tree codes,
and hierarchical grid methods such as ours also lend themselves
well to this approach.
The main difficulty is that in order to keep the total number
of particles manageable, one must essentially run a preliminary
simulation simply to find out where the mass that flows into the
region of interest originated and sample it with a finer
granularity in the initial conditions.
This may lead to problems of contamination by more massive
``background'' particles if the small-scale structure that is not
resolved by the preliminary simulation turns out to have a significant
impact on the dynamics.
Furthermore, the presence of particles of different masses in the
system could lead to spurious mass segregation effects.
For these reasons, we attempted to keep the overall size of the
computational box as small as possible, handling the tidal
fields on larger scales, as well as any mass flows (as long as they
remained moderate) as externally imposed boundary conditions.

This turned out not to be a particularly successful design choice,
and we cannot recommend it to others.
Injecting too many, or too few, particles can have a destabilizing
effect.
The first term on the right hand side of equation~\ref{q:motion}
has, in the usual case $\rhobar\ge 0$, the effect of accelerating
particles towards the boundary of the box. This is balanced by the
second term, which represents the attraction between the gravitating
particles. If more particles leave the box than are injected, the
first term will tend to dominate and cause even more particles to be
ejected;
conversely, if too many particles are added the material will tend to
collapse towards the center.
This may be taken to represent a physical effect: a void is expected
to expand faster 
than the universal average, an over-density more slowly.
But unless the algorithm for replenishing the box with particles is
well thought out, a runaway instability may occur.
In practice we have found it expedient to couple the injection
of particles to the outflow so as to maintain a constant mass within
the system.
We then monitor the cumulative mass of particles so recycled and
compare it with the total mass in the box. If the ratio becomes too
large (greater than 10\% or so), we take it as an indication that
one needs to simulate a larger volume of space.
As the volume that needs to be simulated grows larger, our original
motivation for using isolated boundary conditions becomes weaker.
It turns out that the method we described in this paragraph works
much better if the system is located in a region of lower density
than the cosmic mean, as the mass fraction that undergoes reinjection
is smaller in that case.
This allows us to simulate a number of interesting systems, but
is not suitable for statistical studies involving a random selection
of initial conditions.

Momentum, angular momentum, and energy can be carried by the particles
that flow in and out of the system and by direct interaction with the
external potential $\Phi_x(\vecx;t)$ and
the uniform density we impose in the space that surrounds the particle
region.
The latter unfortunately has the cubic symmetry of the computational
box rather than the more convenient spherical symmetry that would be
required for an exact cancellation of the induced tides.
Consequently, these tides are always present except in the pure
non-cosmological case with $a$ a constant, where the background
density vanishes.
Whether this is a serious drawback depends on the physical problem
being studied. If necessary, one can enlarge the computational box
(this may be required in any case to keep mass flows under check),
include a compensating term in $\Phi_x$, or both.

\subsection{Introducing subgrids}

The idea of adding subgrids in regions where better spatial resolution
is required is not a new one.
Past approaches have differed on whether to add mass resolution at
the same time by splitting the particles into a larger number of less
massive ones, on how to minimize errors at the interface between the
finer and the coarser grid, and on whether the solution on the coarser
grid should be modified to take into account the results from the
finer grid.

\subsubsection{Subgrids and particles}

Unlike Villumsen \nocite{Vi89} (1989) and Splinter \nocite{Sp96}
(1996), we do not automatically introduce a new set of less massive
particles on each subgrid.
We take the view that our initial mass granularity already matches the
resolution we wish to achieve, and simply increase the force
resolution (by adding subgrids) when and where the particle density is
high enough to make this permissible and worthwhile.
This allows the decision of where to place subgrids to be made on a
step by step basis, and spares us the need to perform a first
simulation without subgrids to find out where particles of smaller
mass should be placed in the initial conditions.
An additional advantage of having a single set of particles of equal
mass is that we need not worry about mass segregation effects.
A drawback is that maintaining equivalent resolution within a larger
computational volume requires more particles.
Of course our code does support multiple particle masses, and as will
become clear below the criteria for subgrid placement can be tuned to
favor tracking the lower-mass particles; we may therefore decide to
experiment with particles of different masses in future.

In our scheme the various grids are merely devices, introduced
independently on each integration step, to compute inter-particle
forces.
In this respect we are closer in spirit to Couchman's \APPPM{} than
to most other multiple-grid approaches.
Ours is effectively (in the terminology of ANC) a
Particle-Multiple-Mesh (\PMM) scheme.
This distinction will have important consequences below.
Since there is only one set of particles that exist independently of
any subgrids, the equivalent of a back-reaction from the subgrid
solution to the parent grid is automatically included: it is mediated
by the particles themselves.

We typically decrease the grid spacing by a factor of two for each
additional level of subgrids.
Each subgrid thus covers a substantial number of cells of its
parent grid.
Other integral refinement factors are allowed by our formulation
(subject to the condition that the boundaries of any subgrid must
coincide with cell boundaries of the parent grid) but become
increasingly inefficient in terms of volume coverage and we have not
tried them in practice.
They would also exacerbate any ``ringing'' at the interface between
a subgrid and its parent grid; since this is one of the most delicate
aspects of any hierarchical grid scheme, it is probably not advisable
to use refinement factors larger than two under any circumstances.

A particle passing from a region of low resolution to one of higher
resolution effectively undergoes a sudden change in its spatial
extent.
(In \PM{} codes, particles are best thought of as extending over about
one grid cell or more, depending on the shape of the interaction law
and on the charge assignment scheme.)
In its current state, our code takes no particular precautions against
any transients that may result.
Our comparison tests between runs that use adaptive subgridding and
runs with uniform resolution show that such transients are not
important enough to produce significant discrepancies in the results.
This is fortunate, since otherwise we would have had to associate a
time-dependent smoothing length with every particle, which would
complicate the method.
That the smoothing length would have to be associated with the
particle rather than with the location relative to a subgrid follows
both from our choice to treat the particles as primary and the grids
as auxiliary objects and from the fact that new subgrids may be added,
removed or relocated to follow the flow at every step.

\subsubsection{Forces on subgrids}

Our strategy for computing the forces between subgrid particles
differs slightly from that of Villumsen.
His approach was to recalculate the potential on the entire parent
grid after having set the density in the volume occupied by the
subgrid to a constant value equal to the mean density within the
subgrid,
and use this solution as the tidal field on the subgrid particles from
the rest of the system.
The forces between subgrid particles were then computed in the usual
way by solving for the potential on the subgrid.
Instead, we first compute forces for all particles on the parent grid,
then generate the density array on the subgrid (whose nodes must be
aligned with those of the parent grid) as well as on a coarser grid
that has the same spacing as the parent grid and nodes in the same
locations but covers only the volume of the subgrid.
Since coordinates on the subgrid are affected by the same expansion
factor $a(t)$ as on the parent grid, we must subtract the same
$\rhobar$.
We solve for the forces on all subgrid particles from both these
density arrays, and add the difference to the forces we had previously
calculated on the parent grid.
This avoids double counting and means in effect that pairwise forces
between particles that both lie in the subgrid are always evaluated at
the higher of the two resolutions.
The coarse FFT on the subgrid involves far fewer grid points,
so our approach is more efficient than Villumsen's.
It is also more flexible, in that we only need to store the
accumulated force for every particle and the density potential for a
single subgrid at a time, independently of the number and nesting
depth of subgrids.
The cost of actually storing a force vector for every particle may
seem high when one knows that in a traditional \PM{} code with
leapfrog time stepping the
particle accelerations can be interpolated when the velocities are
updated and need not be kept afterwards; but saving them allows us
to use them to adjust the time step and to compute subgrid corrections
to the energy conservation test.
On the other hand, we could have supported independent time steps had
we chosen to save the potential on each subgrid instead.
This, however, would have caused the storage requirements to scale
with the subgrid nesting depth. 

Let $S$ be the set of particles in the subgrid, $C$ its complement.
Schematically, the acceleration on particle $i$ is computed as
follows.
For $i\in C$,
\begin{mathletters}
\begin{equation}
\label{q:acc-pgr-1}
\vecF = \sum_{j\in C\cup S} m_j \nabla_1 V_P (\vecx_i,\vecx_j).
\end{equation}
For $i\in S$,
\begin{eqnarray}
\label{q:acc-sgr-1}
\vecF  = \quad \sum_{j\in C} m_j \nabla_1 V_P (\vecx_i,\vecx_j) &
+ & \sum_{j\in S} m_j \nabla_1 V_P (\vecx_i,\vecx_j) \nonumber \\
 + \sum_{j\in S} m_j \nabla_1 V_S (\vecx_i,\vecx_j) &
- & \sum_{j\in S} m_j \nabla_1 V_{P'} (\vecx_i,\vecx_j)
\end{eqnarray}
\end{mathletters}%
where $V_P$ is the approximation to the Coulomb potential
on the parent grid, $V_S$ the approximation on the subgrid,
and $V_{P'}$ that on the subset of the parent grid that covers the
subgrid.
(In reality, the sums are evaluated by FFT and include the $\rhobar$ term.)
$V_P=V_{P'}$ to a very good degree, provided only that the grid nodes
are in the same locations. (This is an essential requirement:
we have found that violating it has deleterious effects on the
solution.)
The two rightmost terms in the display of equation~(\ref{q:acc-sgr-1})
therefore cancel each other.
Note that we add together force contributions from the various grids
rather than the potentials; this allows the arbitrary additive
constant in the potential to differ from grid to grid without
affecting our solution.

\subsubsection{Boundary conditions on subgrids}

We impose isolated boundary conditions on the subgrid, with an
external background density ($\rhobar$) 
equal to that used on the parent grid.
It may seem more appropriate to use the mean density within the
subgrid, or---better yet---the mean density in the immediate vicinity
of the subgrid;
but one should note that some ringing is to be expected no matter what
the exact scheme adopted since the density at the boundary of the
subgrid always has some short-wavelength component that is resolved on
the subgrid but not on the parent grid.
The usual approach in codes of this type is to try to limit the
ringing by introducing 
around the subgrid a buffer region, in which particles contribute to
the potential but are not affected by it, and/or a transition region
in which the potentials of the parent grid and of the subgrid are
blended linearly.
The transition region implements the variable smoothing length we
alluded to at the end of the previous subsection.
As already noted, it is of limited usefulness in our case since it
doesn't guard against transients induced by the sudden activation of a
new subgrid.
The possible motivation for introducing a buffer region can be
understood by considering the case of a particle $P_0$ located on the
subgrid but near its edge, and subject to the forces of two nearby
equidistant particles $P_S$, also on the subgrid, and $P_B$ just
outside it.
In the exact solution, $P_S$ and $P_B$ exert forces of equal magnitude
on $P_0$. Without a buffer region, the calculated force from $P_S$ is
stronger. This can create spurious small-wavelength perturbations as
the pair $(P_0,P_S)$ tends to collapse on its own rather than as a
triplet $(P_S,P_0,P_B)$.
The introduction of a buffer region would allow the balance to be
preserved between the attractions of $P_S$ and $P_B$.

Our implementation of the buffer region is as follows. 
Extending the notation introduced in the previous subsection, we have
split $C$ into the buffer region $B$ and a distant exterior $X$.
Equation~\ref{q:acc-sgr-1} is modified in that the sum over $j\in C$
becomes a sum over $j\in X$ and the sums over $j\in S$ become sums
over $j\in (S\cup B)$.
We normally set the thickness of this buffer region to be equivalent to
three cells of the parent grid: the grid-based force does not differ
significantly from the exact Coulomb force at separations this large,
so that $V_S \simeq V_P$.
Since the thickness of the buffer region is fixed in units of the
spacing of the parent grid, large refinement factors result in the
buffer region ``eating away'' most of the volume of the subgrid.
This is one of the reasons why we only experimented with a refinement
factor of 2 at each level.

\subsubsection{Adjacent subgrids}

An essential feature of this buffer region is that it enables us to use
multiple adjacent subgrids to cover extended structures that are
worthy of higher resolution but do not fit within a single grid.
This could occur either because of their size or because they are
located too close to other high-density objects already covered by
their own subgrids. (We do not allow arbitrary overlap of subgrids,
for both simplicity and efficiency.)
A problem that might arise when the boundary between two adjacent
subgrids cuts through the middle of a high density structure is that
the force between two nearby particles separated by the boundary would
be calculated with the (poorer) resolution of the parent grid.
Thanks to the buffer region around each subgrid, this difficulty is avoided:
the force exerted by either particle on the other will be
computed at subgrid resolution by virtue of each particle
lying in the buffer region of the other's subgrid.

In the overlap region between adjacent subgrids, it matters little
whether the border region is implemented as described above or in the
following alternative way, by subjecting particles in $B$ to the
subgrid forces without including them in the subgrid potential. This
corresponds to applying equation~\ref{q:acc-pgr-1} to particles $i\in
X$ only, and equation~\ref{q:acc-sgr-1} to particles $i\in (S\cup B)$.
Our first implementation used the latter approach, and comparison
tests show only minute differences between the results of both variants.

A difficulty arises wherever the buffer region has finite width and
does not overlap with an adjacent subgrid.
Then the balancing of the forces by $P_S$ and $P_B$ over $P_0$ comes
at the cost of a violation of Newton's second law: the force exerted
by $P_0$ on $P_B$, being computed on the parent grid, does not balance
the force by $P_B$ on $P_0$.
The net result is that $P_B$ will be accelerated outwards, away from
the subgrid.
(If the alternative implementation of the buffer region is adopted,
the net effect has opposite sign and tends to push the particles
inwards.)
The consequences for the orbit of a bound clump can be spectacular, as
the following worst-case example, illustrated by figure~\ref{f:tf4g}, shows.
We launch a 4096-particle realization of a truncated isothermal sphere
(with a half-mass radius of 0.02 units, corresponding roughly to half
the cell width on the top grid) with a mean velocity of 0.1 units
towards a region covered by a fixed subgrid (the inner bounds of which
are indicated by dotted lines in the figure) with  refinement factor
of two and a buffer region of width 0 (solid curve) or 2 (dashed
curve) parent grid cells.
For this example we adopted the alternative implementation of the
buffer region, which causes the particles to be accelerated inwards.
The initial velocity is along the $x$ axis, perpendicular to the
faces of the subgrid. The figure shows the $x$ coordinate of the
center of mass of the sphere as a function of time $t$.
When the buffer region is suppressed (width 0, solid curve) the clump
enters and exits the subgrid without changes to its bulk velocity.
(The apparent turnaround occurs when the clump reaches the edge of the
computational volume and some particles leave the box. The solid curve
reflects the mean velocity of only those particles that remain in the box.)
The presence of a finite buffer region, by contrast, causes the
clump to accelerate as it enters the subgrid, and to bounce back 
as it tries to exit on the other side.
The reason it doesn't simply lose the momentum it gained when
entering the subgrid is that the momentum change depends on the
difference between the forces calculated on the subgrid and on the
parent grid, and that the passage through the more highly resolved
subgrid has allowed the clump core to relax to a more sharply peaked
density profile, increasing the force mismatch.
The main features of this behavior are quite generic: a more carefully
constructed clump with a larger number of particles that was allowed
to settle into a numerical equilibrium before being launched through
the subgrid underwent similar, if slightly milder, accelerations and
rebounds.

Clearly it is better for us to suppress the buffer region where there
is no adjacent subgrid. The violation of momentum conservation that a
finite border region implies can have a drastic influence on the orbit
of a bound clump.
This would not be much of a concern if we could guarantee that bound
clumps are always entirely covered by subgrids at the same resolution;
the problem does not arise for smooth distributions of matter where
short-range forces don't predominate.
(In section~\ref{s:Gunn-Gott-test} we demonstrate this fact through
a test of smooth infall onto a localized seed mass perturbation.)
In principle, an adaptive algorithm for subgrid placement based on
the density should tend to ensure this. However, one may want to
place additional restrictions on which regions are followed at higher
resolution, in which case the guarantee does not hold and we must take
the precaution of suppressing the buffer region in the absence of an
adjacent subgrid.
Note that the problem is less severe when expanding coordinates are
used, as in cosmological applications, since the peculiar velocity
acquired while crossing a subgrid interface will be damped away.

Why has this same difficulty not been recognized by other authors,
notably ANC and Splinter \nocite{Sp96}(1996)?
The reason may be traced to a fundamental difference in philosophy
between our code and theirs.
The very idea of a particle on a subgrid interacting
symmetrically with a particle on the parent grid is a consequence of
our decision to regard the particles, rather than the density field
on the grids, as primary.
In a scheme where every particle is associated with only one grid,
and particularly when the dynamics of the parent grid particles are
unaffected by the subgrid (the so-called ``one-way interface''
schemes), momentum conservation should be examined separately on each
grid: on the parent grid, no violation is expected, while on the
subgrid the effects of the parent grid particles are treated as an
external field and the buffer and transition regions are used to
smoothly bring any particles exiting the subgrid under the control
of the parent grid field alone, as test particles, rather than
reflecting them back into the subgrid.
Thus, the apparent contradiction between these authors' use of a
buffer region two cells wide and our restriction of the buffer region
to the sole case of adjacent subgrids does not signal an error either
on our part or on theirs, but is merely a manifestation of our
different view of the respective roles of particles and grids.

\subsubsection{Subgrid placement algorithms}

A distinguishing feature of our code is that the activation of
subgrids is entirely automated: the user need only set a few
parameters (maximum depth, and various optional thresholds),
after which the code decides at every step where to place subgrids.
Tuning the criteria for subgrid placement is not an easy task;
our current choices can undoubtedly be improved upon.
Here we shall describe some of the criteria we have implemented.
Not all of them have proved very useful in practice.

Our subgrids have fixed shape and size, and may not overlap.
Consequently, we found it simplest to introduce a uniform tiling of
subgrids covering the entire volume of the parent grid. 
Each subgrid may be either active or inactive.
A subgrid is activated whenever our requirements for subgridding are
satisfied within its volume.
We are free to choose the origin of the tiling independently for every
parent grid at every step.
We make use of this freedom to reduce the number of
active subgrids. Typically we try eight different possible origins,
and adopt one that requires the activation of fewest subgrids.

The first criterion for subgrid activation is that the particle number
density be sufficiently high in at least part of the covered region:
there is no point in the grid ever being much finer than the
smallest distance between particles.
To activate a subgrid, we require that among the corresponding cells
of the parent grid either (a) at least one cell contains more than
$N_0$ particles; or (b) one cell contains at least $N_1$ particles and
its 26 immediate neighbors together contain a further $N_3-N_1$
particles or more.
$N_0$, $N_1$ and $N_3$ are specified as input parameters to the code.

The main reason for considering cubes of $3\times3\times3$ cells as
well as single cells is that one needs to detect the collapse of a
high-density region before it is too advanced: the additional
resolution is already required during the later stages of collapse.
We don't try to detect larger, lower-contrast potentially collapsing
structures on larger scales since for separations larger than about
three cells the forces are computed accurately on the parent grid.

One way of choosing $N_1$ and $N_3$ is the following. Let $\Neff$ be a
measure of the mean number of particles per cell in some larger
region, such as the entire candidate subgrid. (This measure need not be
unbiased; in fact, if the mean particle number density is very low we
found it necessary to artificially increase $\Neff$ by setting it to the
value 1.
Otherwise our subgrid placement criterion would be satisfied far too
easily in low-density regions.)
If the particle counts were distributed according to Poisson
statistics, one would expect a group of $\Ncells$ cells to contain on
average $\Ncells\Neff$ particles, with a standard deviation
$(\Ncells\Neff)^{1/2}$ around that mean.
By setting $N_1=\Neff+N_\sigma \Neff^{1/2}$ and $N_3 = 3^3 \Neff +
N_\sigma (3^3\Neff)^{1/2}$, where $N_\sigma$ is a tunable parameter,
we try to ensure that only particularly significant deviations from
homogeneity trigger subgrid activation.

We have found that this approach works reasonably well during the
early stages of nonlinear evolution, where it causes higher resolution
to be applied to the regions we are most interested in.
However, as the simulation progresses we have needed to
increase $N_\sigma$ in order to limit the proliferation of subgrids,
as our goal was merely to achieve high resolution in a few large peaks
near the center of the computational box.
Clearly the criteria can and should  be tuned according to the needs of
individual applications; no single choice is optimal for all cases.

\subsection{Time step selection}

We follow the usual practice in \PM{} codes of advancing the positions
and velocities of the particles in leapfrog fashion:
\begin{eqnarray}
\label{q:lf-v}
\frac{d\vecx_i}{d\tau} (\tau+\hdtau) & = &
\left(\frac{1-A(\tau)\dtau}{1+A(\tau)\dtau}\right) \frac{d\vecx_i}{d\tau}
(\tau-\hdtau)
+ B(\tau)\dtau \vecf_i (\tau) \\
\label{q:lf-x}
\vecx_i (\tau+\dtau) & = &
\vecx_i (\tau) + \dtau \frac{d\vecx_i}{d\tau}(\tau) .
\end{eqnarray}
An alternative formula applies on a starting step, when both the
positions and the velocities are known at the same time $\tau$:
\begin{eqnarray}
\label{q:lf-start}
\vecx_i (\tau+\hdtau) & = & 
\vecx_i (\tau) + \frac{\dtau}{2} \left(1-\frac{A(\tau)\dtau}{2}\right)
\frac{d\vecx_i}{d\tau} + \frac{\dtau^2}{8} B(\tau) \vecf_i (\tau) .
\end{eqnarray}
Likewise, one can synchronize positions and velocities at the final
step by advancing the velocities from $\tau-\hdtau$ to $\tau+\hdtau$
then updating the positions as
\begin{eqnarray}
\label{q:lf-stop}
\vecx_i (\tau+\hdtau) & = &
\vecx_i (\tau) + \frac{\dtau}{2} \left(1+\frac{A(\tau)\dtau}{2}\right)
\frac{d\vecx_i}{d\tau} - \frac{\dtau^2}{8} B(\tau) \vecf_i (\tau) .
\end{eqnarray}

In large simulations, it is particularly important to avoid taking
more time steps than required to obtain accurate results.
The time step should therefore be allowed to vary.
In the standard leapfrog formulation, 
the cancellation of second-order terms in equation~(\ref{q:lf-v})
depends on $A$ being evaluated at the midpoint of the velocity step.
The cancellation of the second-order term in the formula to update the
positions is less important since the $\vecf_i$ are known; however,
it contributes to lowering the operation count for the method.
Various approaches can be used to change the time step in the middle of
a simulation. One is to use a different midpoint in
equation~(\ref{q:lf-v}) and calculate the corresponding $A(\tau)$.
The other, which is the one we adopted, is to keep $\dtau$ constant
by an appropriate choice of the function $\tau(t)$.
We construct this function and its first and second derivatives,
which are needed to evaluate $A$ and $B$ according to equations
(\ref{q:A}) and~(\ref{q:B}), step by step as the simulation
progresses.
(In practice, we do not allow $d\tau/dt$ to vary too quickly, since
that tends to spoil the results. We shrink $\delta t$ by at most 25\%
on each step, and let it grow even more slowly.
Situations in which this forces a larger $\delta t$ than desired have
fortunately been very rare in our runs.)

Our preferred criterion for determining the time step is the Courant
condition:
\begin{equation}
\label{q:Courant}
\delta t \le \varepsilon \min_j \frac{h_j}{\norm{\vecxd_j} }
\end{equation}
where the index $j$ spans the set of all particles, $h_j$ is the grid
spacing of the finest grid used to compute the forces on particle $j$,
and $\varepsilon <0.5$ is a constant coefficient.
$\vecxd_j$ should be taken as being the mean velocity over the time
step, and is a function of both the velocity before the step is taken
and of the acceleration.
The latter can be important if the initial velocity is
uncharacteristically small, for example when all the particles are
initially at rest.
The equation for the maximum acceptable $\delta t$ for each particle
is actually a quartic. We do not solve it exactly, preferring to
estimate a close lower bound on the solution for the particles with
the tightest $\delta t$ constraints.
This way of taking the accelerations into account fulfills the same
function as other authors' use of the maximum density on a subgrid to
estimate a local dynamical time.

In some cosmological simulations with a large number of particles, we
found that adopting the smallest $\delta t$ found in this fashion was
leading to time steps much shorter than warranted by our physical
understanding of the dynamical time scales involved (based on the
maximum density). This turned out to be due to a small number of high
velocity particles: the tail of the velocity distribution produced by
the violent relaxation of a newly collapsed object, which is naturally
better sampled as the number of particles in the simulation increases.
Since these particles represent a small fraction of the mass in the
typically high-density regions in which they are found, their motion
has almost no impact on the mean gravitational field, which evolves on
much longer time scales.
This led us to modify the time step criterion for these simulations by
pretending that the velocity of each particle before the time step is
negligible and basing the choice of $\delta t$ solely on the
accelerations.
This is not unreasonable for cosmological applications since in the
linear regime the velocities and peculiar accelerations are related
and in virialized clumps they are a better indication of the dynamical
time scale than the velocity of the fastest particle (which is not the
same as the local velocity dispersion).
Moreover, expanding coordinates have the property that peculiar
velocities are damped away and need to be continually regenerated by
the forces.
As a further precaution, we have analyzed the velocities of the
particles at a late stage in a simulation run with this relaxed
time step criterion and found, as intended, that the number of
particles that violated the stricter Courant condition was small.
The relaxed criterion is known to fail, however, for highly ordered
collapse (such as that of a homogeneous sphere). 
In that case the
largest acceleration does not provide a good estimate of the time
scale over which the mean field evolves: based as it is on the more
extended distribution at the beginning of the step, it systematically
overestimates the maximum allowable time increment.
Accordingly, we only resort to the relaxed criterion, with some
reluctance, for runs in expanding coordinates, with large numbers of
particles, and where different mass shells collapse at different
times. Our tests on smaller cosmological runs have shown the results
to be substantially identical, but with the relaxed condition
requiring a much smaller number of steps.

We use a single value of $\delta t$ for all particles on all subgrids
at any given step.
A far better approach in principle, but much more complicated to
implement, would be to adopt multiple time steps and to distinguish
between the time steps used for advancing individual particles and
those for updating the potentials on subgrids.
Independent time steps for the particles would require a way of
estimating the accelerations at positions and times slightly different
from those for which the potential has been calculated.
For this, it may be necessary to calculate the time derivative of the
subgrid potential as well as the potential itself.
In addition to the constraints resulting from the rate of change of
the subgrid density, the frequency with which the subgrid potential is
recalculated can also be affected by particles flowing out of the
region in which forces can be interpolated from the subgrid potential:
in our scheme, such events change the mapping between particles
and subgrids and require recalculation of all the subgrid potentials
involved if the total momentum is to be conserved.
These are our main reasons for adopting a single time step for
all grids.
The other main reason is that a single time step allows us to process
subgrids one by one, accumulating the force contributions on each
particle without needing to store the solutions for the potential on
many nested subgrids simultaneously. Almost any other time step scheme
requires these subgrid potentials to be available simultaneously.

The coefficients in the leapfrog formulae (equations~\ref{q:lf-v}
and~\ref{q:lf-x}) can depend on time.
For good accuracy, their variation in a time step must be small.
This constraint is clearly unrelated to the grid spacing, and must be
imposed separately. It is equivalent to the requirement that
$|H \delta t| \ll 1$.

\subsection{Conservation laws}

Conservation laws provide an important diagnostic of how accurately
the equations of motion have been integrated in a given simulation.
Good conservation of physical invariants is not a sufficient condition
for a good integration, but it is a necessary one.

In the absence of coordinate expansion, subgrids, and external fields,
a \PM{} code such as ours would be expected to conserve total momentum
algebraically, energy a little less well, and angular momentum only in
the limit where the interparticle separation is much larger than the
grid spacing (HE \S 7--6).

\subsubsection{Energy}

Cosmic expansion causes the usual equation of conservation of the
total energy, $E=T+W={\rm constant}$ (where $T$ is the kinetic, $W$
the potential energy), to be replaced with the Layzer-Irvine
(\cite{Ir61}; \cite{La63}) equation, which can be derived from the
well-known property of the Hamiltonian $\Ham$:
\begin{equation}
\label{q:dHam-dt-law}
\frac{d\Ham}{dt} = \frac{\partial\Ham}{\partial t}.
\end{equation}
Computing the canonical momentum from equation~\ref{q:Lag-x},
\begin{equation}
\vecp_i \equiv \frac{\partial\Lag}{\partial \vecxd_i} = m_i a^2
\vecxd_i,
\end{equation}
it follows in the usual way that
\begin{eqnarray}
\Ham & = & \sum_i \vecp_i \cdot \vecxd_i - \Lag \nonumber \\
& = & \half \sum_i \frac{\vsq{\vecp_i}}{m_i a^2}
- \frac{2\pi}{3} \rhobar a^2 \sum_i m_i \vsq{\vecx_i}
- \half \sum_i \sum_{j\ne i} m_i m_j V(\vecx_i,\vecx_j; t)
+ \sum_i m_i \Phi_x(\vecx_i; t).
\end{eqnarray}
We define the kinetic and potential energies as:
\begin{eqnarray}
\label{q:T}
K & \equiv & a^2 T \equiv \half \sum_i \frac{\vsq{\vecp_i}}{m_i a^2}
\\
\label{q:W}
W & \equiv & - \half a \sum_i \sum_{j\ne i} m_i m_j V(\vecx_i,
\vecx_j; t)
- \frac{2\pi}{3} a^2 \ddot{a} \sum_i m_i \vsq{\vecx_i} .
\end{eqnarray}
Then
\begin{equation}
\Ham = a^2 T + a^{-1} W + \sum_i m_i \Phi_x (\vecx_i; t).
\end{equation}
Our exclusion of the external potential $\Phi_x$ from the definition
of~$W$ is somewhat arbitrary. In practice, one wants to separate the
terms associated with the mutual interactions of particles, which do
measure the accuracy of the integrator, from the correction terms
that merely account for known, external sources and sinks of energy.

In applying equation~\ref{q:dHam-dt-law}, one notes that in
\begin{equation}
\frac{\partial a^{-1} W}{\partial t} = -H a^{-1} W 
+ a^{-1} \frac{\partial W}{\partial t}
\end{equation}
the last term does not vanish if $aV$ depends explicitly on time (as
may occur in practice when a new subgrid is activated and the force
resolution increases locally), or if $a^3\rhobar$ is not constant.
For simplicity, we shall assume $a^3\rhobar={\rm constant}$ from now
on, since that is the only case of actual practical interest to us.

Straightforward algebra leads to the two equations:
\begin{equation}
\frac{d}{dt}\left(a^3T+W\right) +Ha^3T +a\sum_i m_i \vecxd_i \cdot
\nabla\Phi_x(\vecx_i;t) + \half \sum_i \sum_{j\ne i} m_i m_j
\frac{\partial aV}{\partial t} = 0
\end{equation}
\begin{equation}
\frac{d}{dt}\left(a^4T+aW\right) -HaW +a^2\sum_i m_i \vecxd_i \cdot
\nabla\Phi_x(\vecx_i;t) + \half a \sum_i \sum_{j\ne i} m_i m_j
\frac{\partial aV}{\partial t} = 0
\end{equation}
which although physically equivalent are evaluated numerically in
slightly different ways when $a(t)$ is not constant.

The corresponding conserved quantities are:
\begin{eqnarray}
C & = &
a^3 T + W + \int H a^3 T \, dt
+ \int a \sum_i m_i \vecxd_i \cdot \nabla \Phi_x (\vecx_i; t)\, dt
\nonumber \\ & &
+ \half \int \sum_i \sum_{j\ne i} m_i m_j \frac{\partial aV}{\partial
t} (\vecx_i, \vecx_j; t) \, dt
\label{q:C}
\end{eqnarray}
\begin{eqnarray}
C' & =  &
a(a^3 T + W) - \int H a W \, dt
+ \int a^2 \sum_i m_i \vecxd_i \cdot \nabla \Phi_x (\vecx_i; t)\, dt
\nonumber \\ & &
+ \half \int a \sum_i \sum_{j\ne i} m_i m_j \frac{\partial aV}{\partial
t} (\vecx_i, \vecx_j; t) \, dt
\label{q:Cprime}
\end{eqnarray}

In proper-length coordinates ($a\equiv 1$), the first integral term
($+\int Ha^3 T\,dt$ or $-\int HaW\,dt$) vanishes and both criteria
reduce to the usual law of conservation of energy.
The last two terms represent respectively the work done by external
fields and time dependence in the interaction law.
One may question the appropriateness of including the latter effect,
since its origin lies in the numerical method used rather than in the
underlying physics. Shouldn't contributions from the
$\partial(aV)/\partial t$ term be counted as violations of energy
conservation by the code?
In reply we point out that much of that term may result from
fluctuations in the zero level of the gravitational potential on
subgrids, which we did not attempt to suppress since they have no
bearing on the computed forces. And in any case the
$\partial(aV)/\partial t$ contribution should be evaluated so that
its magnitude relative to the other terms may be assessed.

It is unrealistic to expect a \PM{} code to conserve $C$ and $C'$
arbitrarily well when gravitational instability causes the individual
terms ($W$, $a^3T$) to grow by orders of magnitude.
Like many other authors before us, we deem $C$ and $C'$ adequately
conserved if their variation is a small fraction (typically
1--3\%) of the change in $W$ or $aW$, respectively.

In our definition of $W$ (equation~\ref{q:W}) we excluded the
self-energy terms $\sum_i m_i^2 V(\vecx_i,\vecx_i;t)$.
But the natural way of calculating the potential in a \PM{} code
effectively includes these terms (the restriction $j\ne i$ can be
ignored since the scheme avoids self-forces); we subtract them
explicitly as part of our energy conservation test.

\subsubsection{Momentum}

We adopted a momentum-conserving scheme for calculating the forces.
The total momentum should therefore be unaffected by particle-particle
interactions on any given grid, and (given the precautions we have
taken in our handling of the buffer region surrounding each subgrid) be
only weakly modified by interactions across the boundary between
adjacent subgrids.
However, it is still affected by all other interactions between
particles and external fields, including the constant-density
background outside the computational box. (This field gives rise to
forces only when isolated boundary conditions are used, which is the
reason why it isn't normally discussed in the literature.)

The canonical momentum $\vecp_i$ of particle~$i$ satisfies
\begin{equation}
\dot\vecp_i = -\frac{\partial\Ham}{\partial\vecx_i} =
\frac{4\pi}{3} \rhobar a^2 m_i \vecx_i
+ m_i \sum_{j\ne i} \nabla_1 V(\vecx_i,\vecx_j; t)
-m_i \nabla \Phi_x(\vecx_i; t).
\end{equation}
Summation over all particles and time integration yield a conserved
quantity
\begin{equation}
\label{q:Pc}
\vecP_c = \sum_i \vecp_i -\frac{4\pi}{3} \int \rhobar a^2 \sum_i m_i
\vecx_i \, dt
- \int \sum_i \sum_{j\ne i} m_i m_j \nabla_1 V (\vecx_i, \vecx_j; t)
\, dt
+ \int \sum_i m_i \nabla\Phi_x (\vecx_i; t) \, dt .
\end{equation}
In the special case when (i) $\Phi_x$ has no spatial dependence, (ii)
$\sum_i m_i \vecx_i \equiv 0$ or $\rhobar \equiv 0$, and (iii)
$\nabla_1 V + \nabla_2 V \equiv 0$, the total momentum is constant.
If one's aim is to measure the deviation of the results from those
expected when $\nabla_1 V + \nabla_2 V \equiv 0$ (such a deviation
can occur in the presence of subgrids), one should naturally omit the
corresponding term from equation~\ref{q:Pc}.

\subsubsection{Angular momentum}

Likewise, the conservation of angular momentum $\vecJ_i \equiv \vecx_i
\times \vecp_i$ amounts to the constancy of
\begin{equation}
\vecJ_c = \sum_i \vecJ_i
- \int \sum_i \sum_{j\ne i} m_i m_j \vecx_i \times \nabla_1 V
(\vecx_i, \vecx_j; t) \, dt
+ \int \sum_i m_i \vecx_i \times \nabla \Phi_x (\vecx_i; t) \, dt.
\end{equation}
Angular momentum is only conserved if (i) 
$(\vecx_i\times\nabla_1+\vecx_j\times\nabla_2)V\equiv 0$ and (ii)
$\Phi_x$ is a function of $\norm{\vecx}$ and~$t$ alone.
The exact Coulomb force law satisfies the first condition, but
grid-based approximations to it don't.
As in the analogous case for the momentum, if the aim is to measure
deviations from the ideal behavior where (i) is satisfied, then one
should compare the magnitude of the corresponding term to $\sum_i \vecJ_i$
itself, and to the magnitude of the external torque term when present.

\section{Tests of the method}
\label{s:tests}

\subsection{Force errors and subgrids}

Our first tests are meant to see how accurately the code calculates
the accelerations of particles given a known mass configuration.
The simplest case is that of the forces due to a single
point mass, which can be compared to the value predicted by Coulomb's
law. Figure~\ref{f:tf1b.dff} 
illustrates this
comparison. It shows the accelerations induced by a single particle at
the center of a $32^3$ grid with isolated boundary conditions. A
single $32^3$ subgrid, twice as fine as the top grid, was centered on
the massive particle.
Two sets of points are readily distinguished, corresponding to test
particles inside and outside the subgrid.
As expected, the acceleration is systematically underestimated since
the effective potential is softer than that of a point mass.
Also as expected, the subgrid extends the range over which the force
is accurate to within a few percent; the accuracy threshold (a little
over 3\% in this test) could be lowered by increasing the size of the
subgrid, the spacing being equal.

Our second test checks that the accelerations computed on a set of
adjacent subgrids agree with those one would obtain on a single,
larger grid with the same spacing as the subgrids.
We compare forces on individual particles randomly chosen to sample
a uniform distribution. The long-range forces, which are adequately
resolved on the parent grid, tend to cancel, so that the system is
dominated by short-range interactions that will be most affected by
the grid refinement.
Our reference forces are from a non-subgridded run with $56^3$ cells
and $56^3$ particles. We compare these to those obtained from three
different runs with only $28^3$ cells on the parent grid, in which
all possible first-level subgrids are activated. The only difference
between the three runs is in the width of the buffer region where
adjacent subgrids overlap.
The results are illustrated in figure~\ref{f:df}. The magnitude of
the force difference is plotted against that of the reference force
for a randomly selected 1\% of the particles (to avoid overcrowding
the plot). 
Panels (a), (b), and (c) correspond respectively to a buffer width 
of 0, 1, and 2 parent grid cells.
The solid diagonal line represents $|\delta\vecF|/|\vecF| = 0.1$;
the last panel also contains a dotted line where
$|\delta\vecF|/|\vecF| = 0.01$.
The results show that when the buffer is sufficiently wide (at least
two cells; we like to use three cells in production runs with larger
grids), the
forces computed by the subgrid method are mostly within 1\% of the
forces from a single grid of equivalent resolution.
Whenever the buffer region is suppressed, an appreciable fraction of the
particles exhibits force errors larger than 10\%.
Here it is essential to understand what we mean by ``errors''.
Figure~\ref{f:df} only compares the forces computed using a set of
abutting subgrids of equal resolution with those from a single, larger
grid with the same spacing, and shows that some overlap is both
required and sufficient to avoid loss of resolution at the boundary
between such abutting subgrids. It does {\em not} compare forces
computed at different grid spacings, and does not address the issue
of convergence to a ``perfect'' solution in the limit of an infinitely
fine grid. We know that at the interface between a subgrid and a
coarser parent grid the accuracy of the forces within the outer two or
three cell layers of the subgrid will be less than that of a uniform
fine grid regardless of the thickness of the buffer region. (It will,
however, be no worse than on the parent grid alone.) We compensate for
this by making the subgrids cover a slightly larger volume than the
region where the higher resolution is required.

\subsection{Collapse of a prolate ellipsoid}

The collapse of a homogeneous self-gravitating pressure-less ellipsoid
initially at rest can be described by a small set of coupled ordinary
differential equations for the lengths of the principal axes, which
can be integrated in a straightforward way (\cite{LMS65}).
This solution provides a convenient test of our code.
The test corresponds to that of collapse to a sheet or filament in
a cosmological code with periodic boundary conditions, but is more
appropriate to our use of isolated boundary conditions. In particular,
the plane-wave test of Efstathiou et al.\nocite{EDFW85} (1985) is only
easy to interpret and compare with a semi-analytic solution when
periodic boundary conditions are used, and these lie outside the scope
of the present article. Another test, in section~\ref{s:voidtest}
below, shows the behavior of our code when integrating the collapse of
a moving sheetlike ridge in the presence of cosmological expansion.

We followed the collapse of a homogeneous ellipsoid with a $2:1:1$
ratio between the lengths of the principal axes. The experiment was
repeated for two different choices of the grid spacing in order to
verify convergence towards the correct solution.
For the higher resolution, we performed two runs, one using a single
grid of $128^3$ cells and one with a top-level grid of only $32^3$
cells but up to two levels of subgrids.
Each subgrid also has only $32^3$ cells; when necessary, multiple
adjacent subgrids were automatically generated by the code to cover
the entire ellipsoid.
The number of particles is the same in all three runs, about $10^5$.

Figure~\ref{f:ts7-overview} shows the short axis $c$ of the ellipsoid
as a function of time $t$. The solid curve represents the analytic
solution, the dot-dash curve the solution computed on the coarse grid,
while the remaining two dashed curves show the results of the two
high-resolution runs. The horizontal dotted line corresponds to the
grid spacing in the high resolution runs; that of the low resolution
grid is four times larger.
Figure~\ref{f:ts7-blowup} presents a blown-up view of the same results
around the time of collapse of the ellipsoid to a spindle.

One sees from these figures that increasing the resolution does indeed
yield a better agreement between the \Nbody{} and analytic solutions.
Furthermore, the results with subgridding are nearly identical to
those obtained with a single grid of equivalent resolution.
In the high-resolution runs, the minimum radius of the spindle is
comparable to the grid spacing. At the lower resolution, it is
significantly less than the grid spacing. We did not investigate the
reasons for this in detail, but it is likely that the convergence is
also affected by the number of particles used to represent the
ellipsoid. 
In these tests we have kept this number constant.

The two runs without subgrids also used a small, constant value of
the time step $\delta t=0.001$. Collapse occurs after 1524 and 1644
steps respectively. The subgridded run, by contrast, used the Courant
condition to adjust the time step; only 125 steps were required to
reach the point of collapse.
The good agreement between the high-resolution results obtained with
both choices of time step confirms the validity of our implementation
of adaptive time steps.
We have also repeated the low-resolution run using our adaptive time
step scheme; on figure~\ref{f:ts7-blowup} the results would be
indistinguishable from those obtained with a fixed time step.

\subsection{Secondary infall across a subgrid boundary}
\label{s:Gunn-Gott-test}

An interesting test of the behavior of mass flows into a subgrid
is the case of accretion from a uniform-density background onto
a point mass (or any other compact, spherically symmetric density
profile).
A regular lattice of $56^3$ particles with zero initial peculiar velocity
was laid down in a computational box of unit side to be evolved with
isolated boundary conditions in an $\Omega=1$, $\Lambda=0$ cosmological model.
We added at the center of the box a single particle of mass
$m=(4\pi/3)\rho_b \hat x_i^3 \hat\delta_i$, where $\rho_b$ is the
mean density of the other particles and of the external background,
$\hat x_i=0.1$, and $\hat\delta_i=1$. 
We followed the collapse to a time $t_f=1500t_i$, corresponding to
an expansion factor $a_f/a_i = (t_f/t_i)^{2/3} = 131$.

A scaling solution is available (\cite{Go75}; \cite{Gu77};
\cite{FG84}; \cite{Be85}): 
$\rho(r) \propto r^{-9/4}$.
This simple law is expected to hold only for those shells that
enclose a mass much larger than that of the initial seed and that have
evolved for a few crossing times after their initial turnaround.
The crossing time for a shell is of the same order as its turnaround
time, and its proper radius is a factor $f\sim 0.5$ times the
turnaround radius. 
Using a linear theory approximation (in which the
density perturbation grows in place) until the turnaround at a mean
enclosed overdensity $\bar\delta\simeq 1$, we find that for a given
initial enclosed overdensity $\bar\delta_i$ the turnaround occurs at
$a=a_i \bar\delta_i^{-1}$, 
the subsequent virialization at $a\sim 2a_i \bar\delta_i^{-1}$, 
and the final comoving radius of the shell is 
$fx_i (a_i/a_f) \bar\delta_i^{-1}$, 
where $x_i$ is the initial comoving radius.
Requiring $\bar\delta_i<0.1$, only shells with
$x_i>\hat x_i (\hat\delta_i/0.1)^{1/3} = 0.1\times10^{1/3}\simeq
0.215$ 
are expected to show self-similar
behavior, and then only for $a/a_i\gtrsim 20$.
One can also evaluate Bertschinger's (1985)\nocite{Be85}
equation~(2.7) with $R_i=\hat x_i$, $\delta_i=\hat\delta_i$, $\tau=1500$
(these values lead to $r_{\rm ta}(t_f)=0.237 a(t_f)$)
and make direct use of his numerically determined similarity solution.

A twist of our simulations is that
shells with $x_i>0.5$ are prevented from falling in by the fact that
they are not entirely included within the simulation volume.
Accordingly, the accretion will be starved for $\bar\delta_i\lesssim 10^{-2}$,
and the last complete shell should virialize at $a/a_i\simeq 200$.

Our main reason for performing this test was to compare the solutions
obtained with various treatments of the border region around a
subgrid. 
We performed four runs: one with a uniform $64^3$ grid, the other three
with a $32^3$ grid and a nested $32^3$ subgrid with a linear
refinement factor of 2. (After subtracting edge cells as discussed in
\S~\ref{s:bc}, the computational box was covered respectively by
$56^3$ and $28^3$ cells.)
In one run the border region was suppressed,
as suggested by our tests with a bound clump crossing the boundary,
while the other two both used a border width of two parent grid cells,
once with the particles in the border region contributing to the
subgrid potential without feeling its effects and once with the
border particles being subjected to the subgrid potential without
contributing to it.
In the last two cases, the net effect of the force asymmetries on
overdense regions results in an acceleration pointing respectively 
away from the subgrid and towards it.

Figure~\ref{f:tf6d-rho} shows the logarithmic density profiles for the
runs we just described.
The differences between the results of
all these runs are minute; we find it difficult to ascribe any
significance to such small deviations.
This is reassuring: it suggests that the exact way in which subgrid
borders are treated does not unduly affect the profiles of collapsed
peaks.
The dotted line connects open triangles that correspond to the values
given by Bertschinger (1985)\nocite{Be85} in his Table~4. At small
radii, his solution tends towards the expected $\rho\propto r^{-9/4}$,
and ours is in good agreement with his. The first few caustics can
easily be identified.

\subsection{Expanding spherical void with compensating ridge}
\label{s:voidtest}

Peebles \nocite{Pe87}(1987; see also \cite{P.89}, hereafter PMHJ) has
proposed and used the following test for cosmological \PM{} codes:
integrating the evolution of a spherically symmetric under-dense region
surrounding by a compensating over-dense shell. 
The interior of such a void expands 
faster than the universe as a whole, and in so doing compresses the
surrounding shell, making its profile higher and narrower.
The evolution can be computed analytically until the time at which the
first density cusp appears. For Peebles' profile,
\begin{equation}
\delta(r) = -\delta_0 \left( 1 - \frac{2}{3}
\frac{r^2}{r_0^2} \right) e^{-r^2/r_0^2}
\end{equation}
with $\delta_0 = 0.1$, this occurs at $a/a_i \simeq 120$ in a flat
$\Omega=1$ universe. (Here $a_i$ is the initial value of the expansion
factor $a$.)
We integrated such a void, realized with $64^3$ particles, on a $32^3$
grid and compared our results to the analytic prediction at
$a/a_i=60$.
Figure~\ref{f:ts6aa} can be directly compared to the corresponding
figure~4 of~PMHJ.
A significant difference between our code and theirs lies in their use
of a staggered grid for the force (\cite{Me86}).
The staggered grid increases the spatial resolution twofold (which is
why these authors used it) at the cost of inducing self-forces on
the particles (for which reason we did not follow their example).
Because of this difference, our results obtained with $64^3$ particles
on a $64^3$ grid correspond to their results with the same number of
particles on a $32^3$ grid.
A comparison reveals differences of detail, particularly in the
structure of the outer parts of the shell, but our results match
the analytic solution as closely as theirs.

We take this occasion to illustrate the energy conservation properties
of the code. Figure~\ref{f:ts6aa-econs} shows the evolution with time
of the conserved quantities $C$ and~$C'$ of equations~\ref{q:C}
and~\ref{q:Cprime} (top), as well as that of the ratio $|\Delta
C|/|\Delta W|$ of the changes in $C$ and in the total potential energy
$W$ from the start of the run (bottom). Apart from an initial
transient due to rounding in the print-out from which the plot was
made, an examination of the ratio shows the energy to be conserved to
about 1--2\% accuracy throughout the run.

\section{Conclusions}
\label{s:conclusions}

We have developed a useful and versatile tool for the simulation of
collisionless systems of gravitating particles.
Our code is particularly suited to situations that call for both
fine-grained mass resolution everywhere and high spatial resolution in
selected regions of high density the exact locations of which need not
be known in advance.
Our method is a development of the well-known Particle-Mesh technique.
Tests indicate that in comparable conditions our code performs
substantially like similar codes described in the published
literature.
When subgrids are introduced to increase the spatial resolution
locally, the results in the regions covered by the subgrids are in
very good agreement with those of a conventional \PM{} code with a
single, finer grid.
If the number of subgridded regions is sufficiently small, our
approach requires less computing time and less memory than the
equivalent single-grid approach.
Furthermore, by not increasing the resolution in regions where the
particle number density is low, we avoid making the behavior of the
code collisional, which would be undesirable for applications to
collisionless systems.

A significant advantage of \PM{} (and its variants \PPPM{}, \APPPM{},
etc.) for cosmological applications is that comoving coordinates and
periodic boundary conditions can be supported in a natural way.
However, periodic boundary conditions are not always a physically
appropriate approximation.
In particular, if one's interest is in systems not much smaller than
the size of the computational cube, periodic boundary conditions
introduce a much stronger coupling between the external tidal field
and the internal dynamics of the system than one would expect to occur
in the aperiodic real universe.
Periodic boundary conditions are only appropriate when the individual
structures of interest are much smaller than the computational volume,
in which case the tides due to periodic images are small;
but then the relatively small dynamic range of conventional \PM{}
methods severely limits the resolution that can be achieved.
By introducing hierarchical subgrids, we are able to extend that
dynamic range; however, it is also tempting to construct the
simulation in such a way that the system of interest fills as much of
the computational box as possible.
This is what led us to implement isolated boundary conditions.
In principle, the additional cost of imposing isolated boundary
conditions (which is relatively small when hierarchical grids are
used since isolated conditions have to be imposed on the subgrids in
any case) is offset by the greater freedom one has to apply an
arbitrary time-dependent external tidal field and to adopt a system of
expanding coordinates that matches the evolution of the simulated
object without necessarily coinciding with the expansion of the global
cosmological model.
Of course our method can also be used in a more conventional way,
with periodic boundary conditions on the top grid; in fact, numerous
simplifications occur when this is done.
In addition, our code should be well-suited to non-cosmological
dynamical simulations, \eg, of brief interactions between already
formed galaxies, star clusters, and so on.

Our approach also has a few limitations.
The most important is shared by all particle simulation methods:
it is generally impossible to improve the spatial resolution without
a corresponding refinement of the time resolution: more time steps
need to be taken.
Other difficulties of note are the non-radial nature of forces at
small separations (which could be cured, at some cost in computing
time, by softening the interaction law and increasing the depth of
subgridding to compensate for this softening), the complexity and
difficulty of tuning the subgrid placement algorithm, and the fact
that subgrids of fixed shape will almost inevitably also cover some
regions where the particle number density is low.
The behavior of the code can become collisional in such regions; this
should not affect the computed structure of high-density
condensations, but may well be a source of high-velocity particles
that lead to shorter time steps according to the Courant condition.
An enhancement to our current code could be to refrain from applying
the refined forces to such low-density cells.
Alternatively, it could prove useful to introduce individual
time-varying softening lengths associated with individual particles,
to represent the fact that particles in this method are to be thought
of as possessing a finite extent that depends on the local number
density.
Another desirable improvement is the adoption of different time steps
at different levels of grid refinement; the savings that can be
achieved from this depend on the detailed structure of the tree of
subgrids, and will be greatest for deep trees with most of their
branches at the finer resolution levels.

\acknowledgments

This work was partially supported by grants NSF-AST-86-57647,
NSF-AST-91-19475, and NASA-NAGW-2224.
Some calculations were conducted using the resources of the Cornell
Theory Center, which receives major funding from the U.S. National Science
Foundation and the State of New York, with additional support from the
Advanced Research Projects Agency, the National Center for Research
Resources at the National Institutes of Health, IBM Corporation, and
other members of the center's Corporate Partnership Program.
We thank Adrian Melott, Richard James, and the referee Randall Splinter
for interesting and valuable remarks on an earlier version of this paper.


\newpage
\figcaption[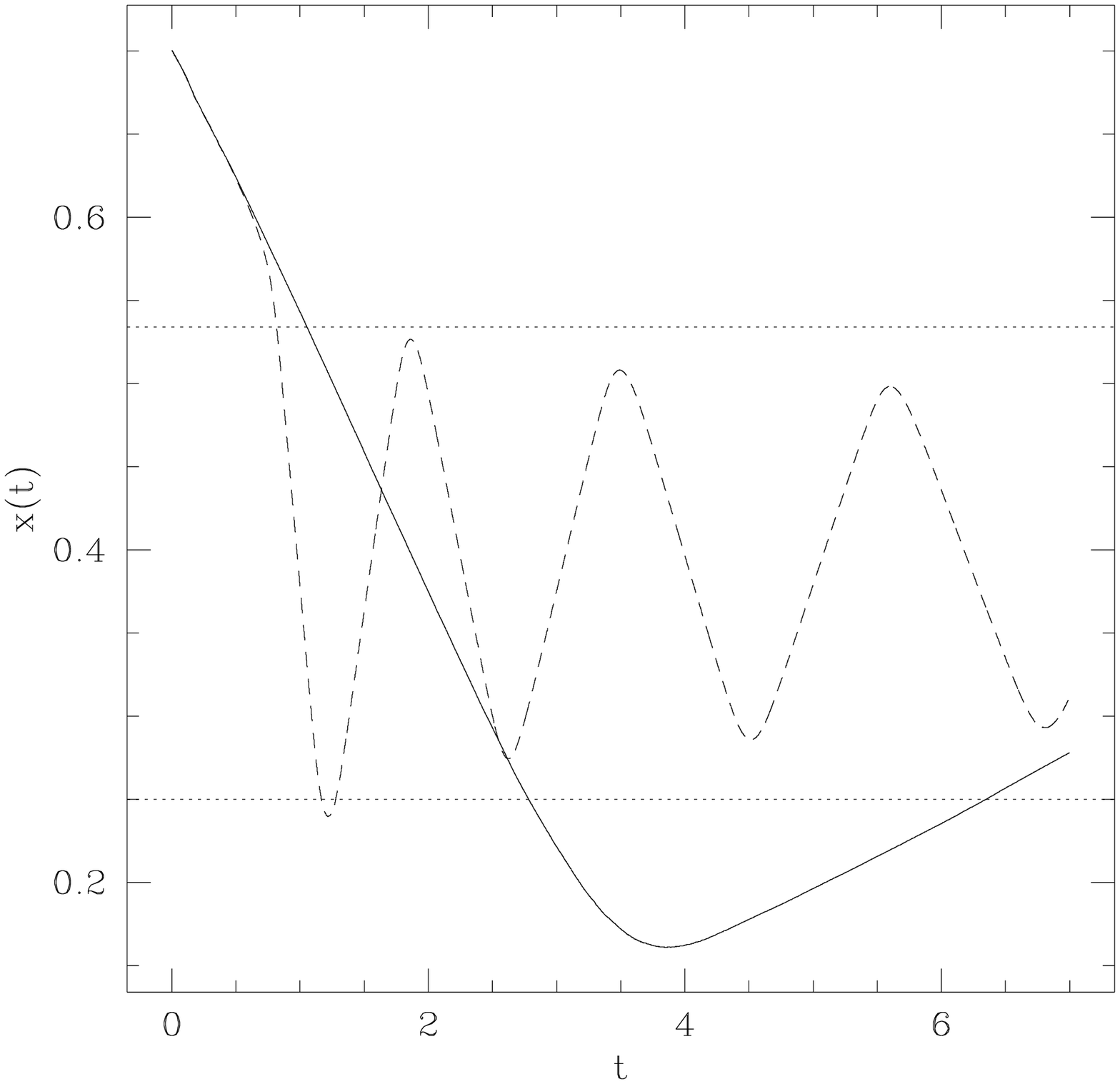]{Trajectory of a bound clump crossing a fixed
subgrid at normal incidence, (dashed curve) with and (solid curve)
without using a buffer region two cells wide around the subgrid.
The figure shows the projected position $x$ of the center of mass
as a function of time $t$. The horizontal dotted lines show the edges
of the subgrid proper; the border region, when used, extends outwards
by 0.07 units on both sides.
The use of a buffer region is seen to produce significant unphysical
changes in the total momentum of the clump.
\label{f:tf4g}}

\figcaption[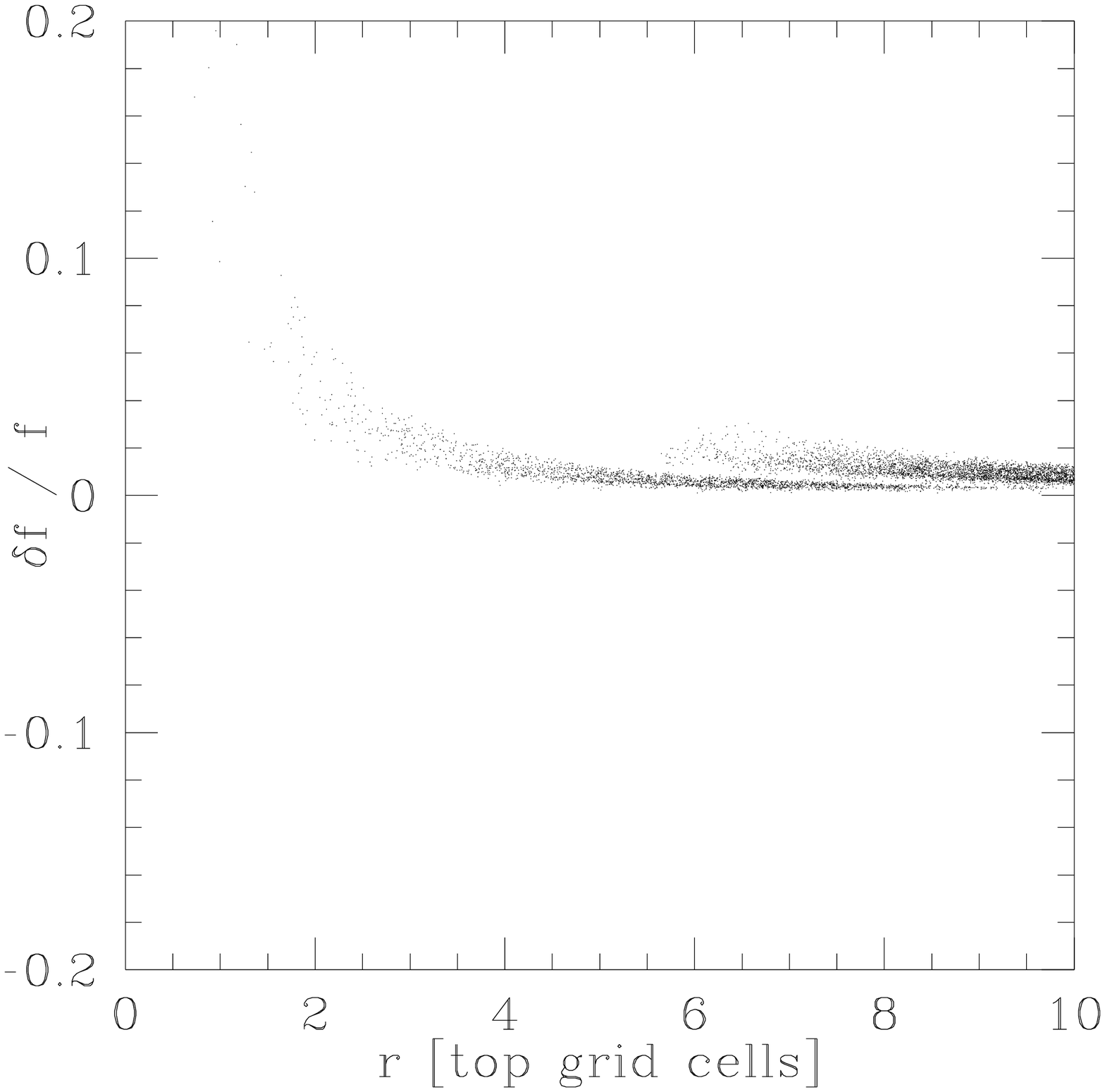]{Relative error in the acceleration induced by a
single massive particle on a number of massless test particles at
various distances. The massive particle lies at the center of a
subgrid whose spacing is half that of the top grid. The abscissa is
measured in units of the top grid spacing; the ordinate is the
relative error in the acceleration compared to the Coulomb formula.
The error only reaches 10\% for separations smaller than $1$--$1.5$
cells (depending on the orientation of the separation vector); it is
less than 4\% outside the subgrid, a number that would be lower still
if larger grid sizes had been used (this test was done with $32^3$
nodes per grid).
The two groups of points correspond respectively to test particles
inside and outside the subgrid.
\label{f:tf1b.dff}}

\figcaption[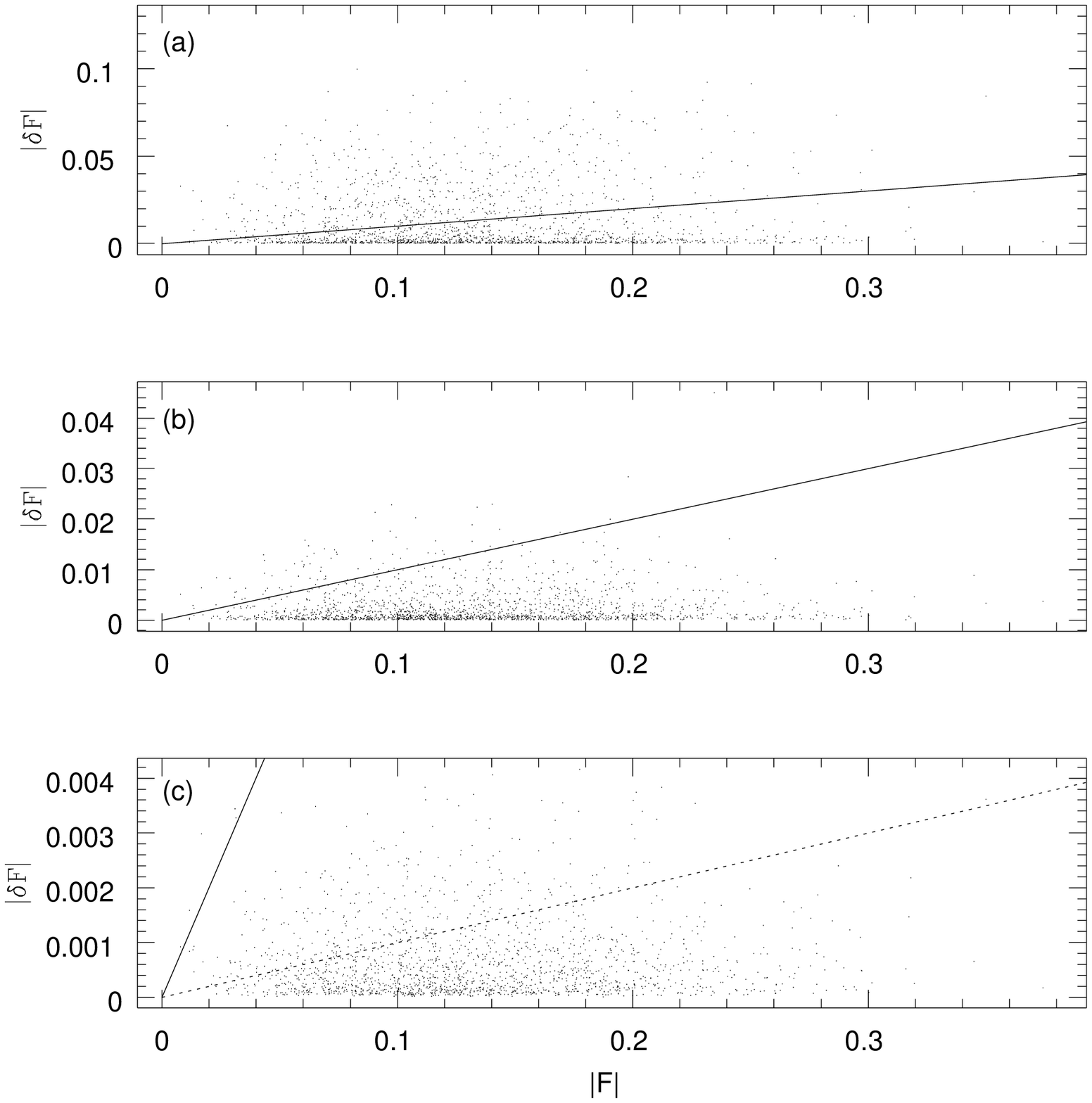]{Comparison of the forces from the subgridding
procedure with those from an equivalent single-grid \PM{} code.
The abscissa is the magnitude of the reference force, the ordinate
the magnitude of the force difference (computed as a three-dimensional
vector). Each point represents one particle; only a random selection
of about 2000 particles is shown.
The force differences depend on the width of the border region
for overlap between adjacent subgrids. Panel (a) corresponds to a
null width, panel (b) to a width of one cell, panel (c) to a width of
two cells.
The solid diagonal straight line shows where the force difference
amounts to 10\% of the force. In panel (c), a dotted line where the
force difference is 1\% is also plotted.
One sees that when the border width is sufficiently large, the subgrid
forces are substantially equivalent to those that would obtain from a
single larger grid with the same spacing.
\label{f:df}}

\figcaption[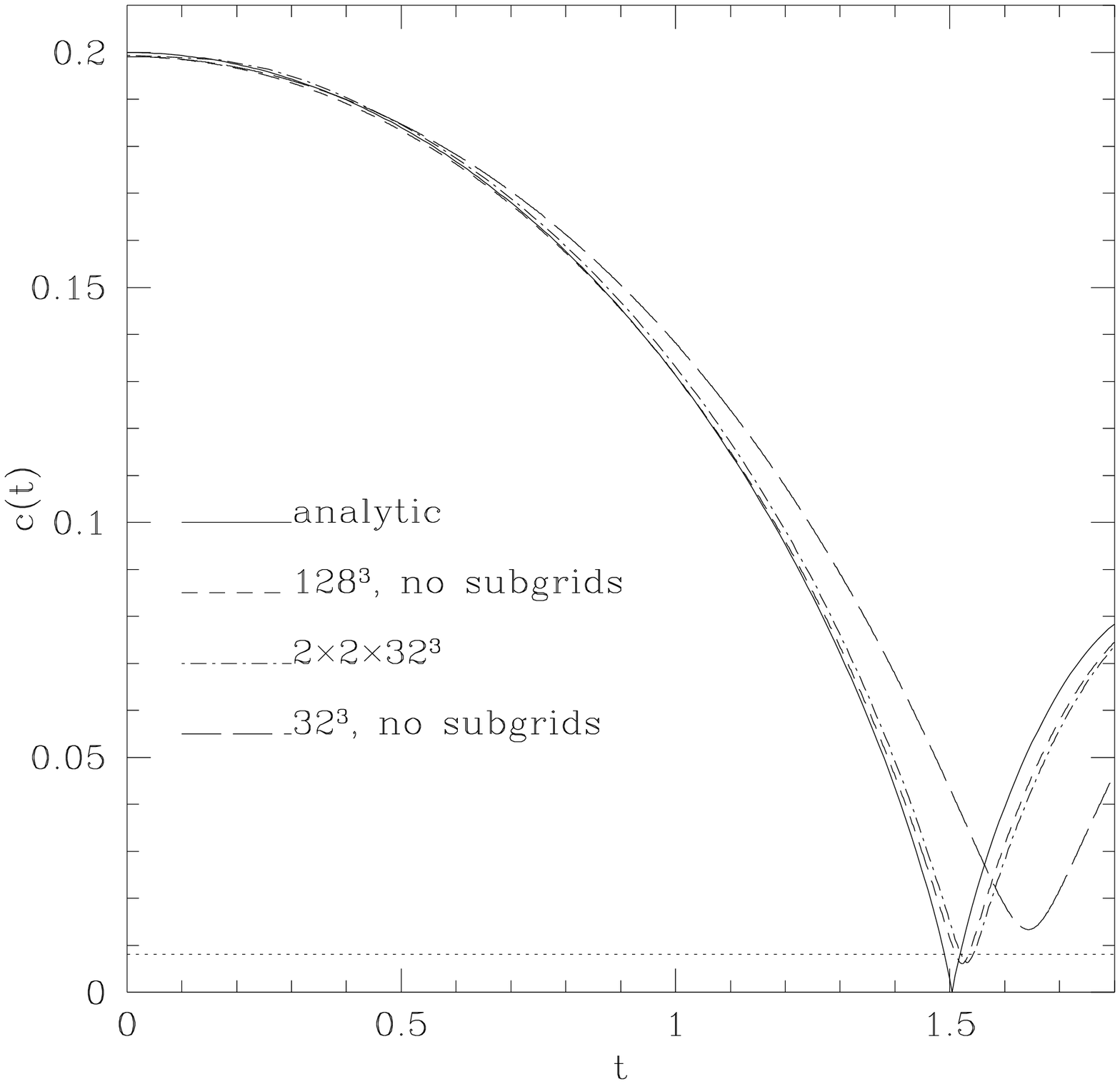]{Minor axis $b=c$ of a collapsing prolate
ellipsoid as a function of time. Solid curve: analytic
solution. Long dashes: low-resolution ($32^3$ grid) \PM{}
simulation. Short dashes: high-resolution simulation ($128^3$ grid, no
subgrids). Dot-dash curve: simulation with $32^3$ base grids and two
levels of subgrids, equivalent to the high-resolution simulation.
\label{f:ts7-overview}}

\figcaption[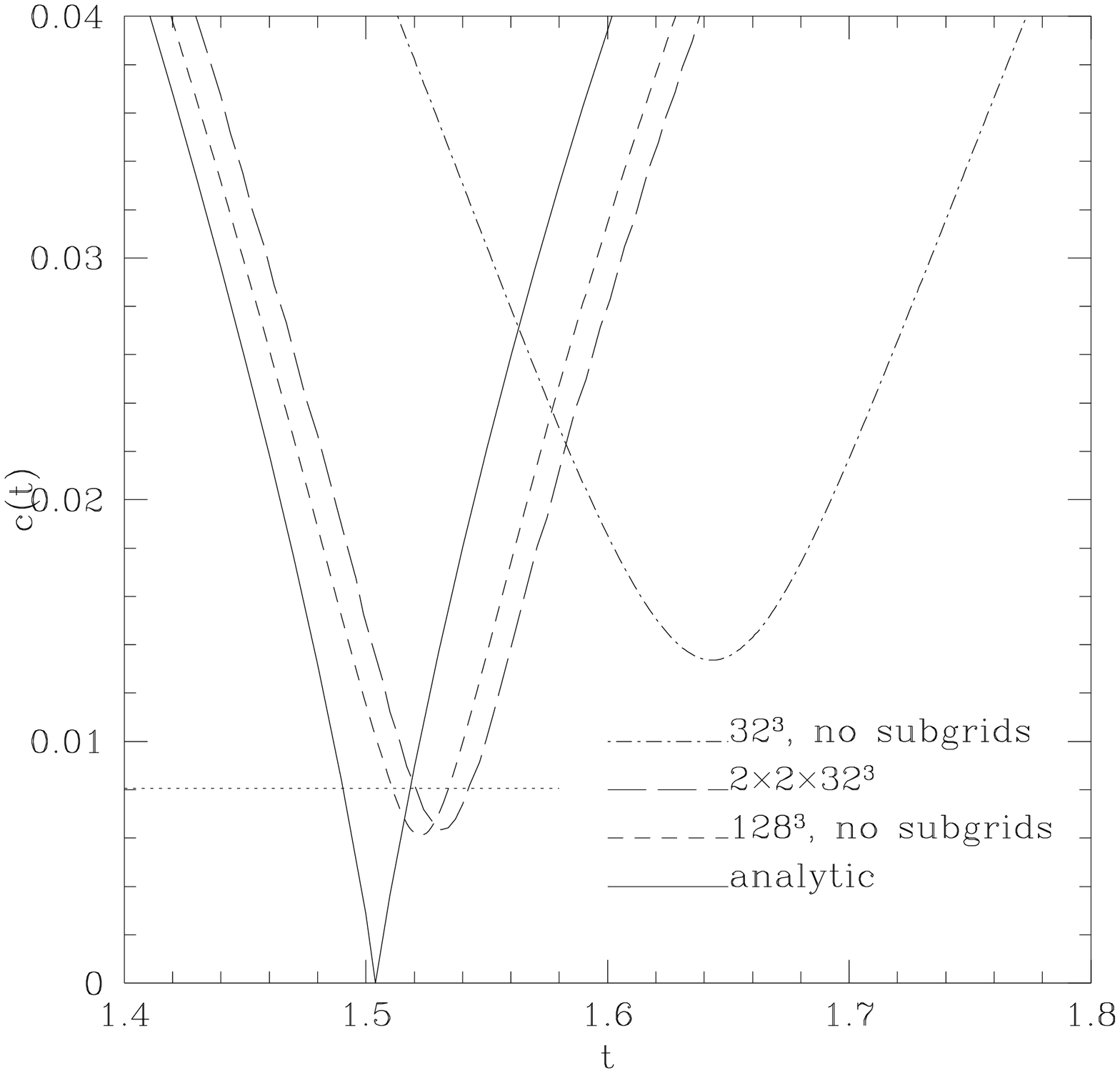]{Enlarged view of figure~\protect\ref{f:ts7-overview}.
\label{f:ts7-blowup}}

\figcaption[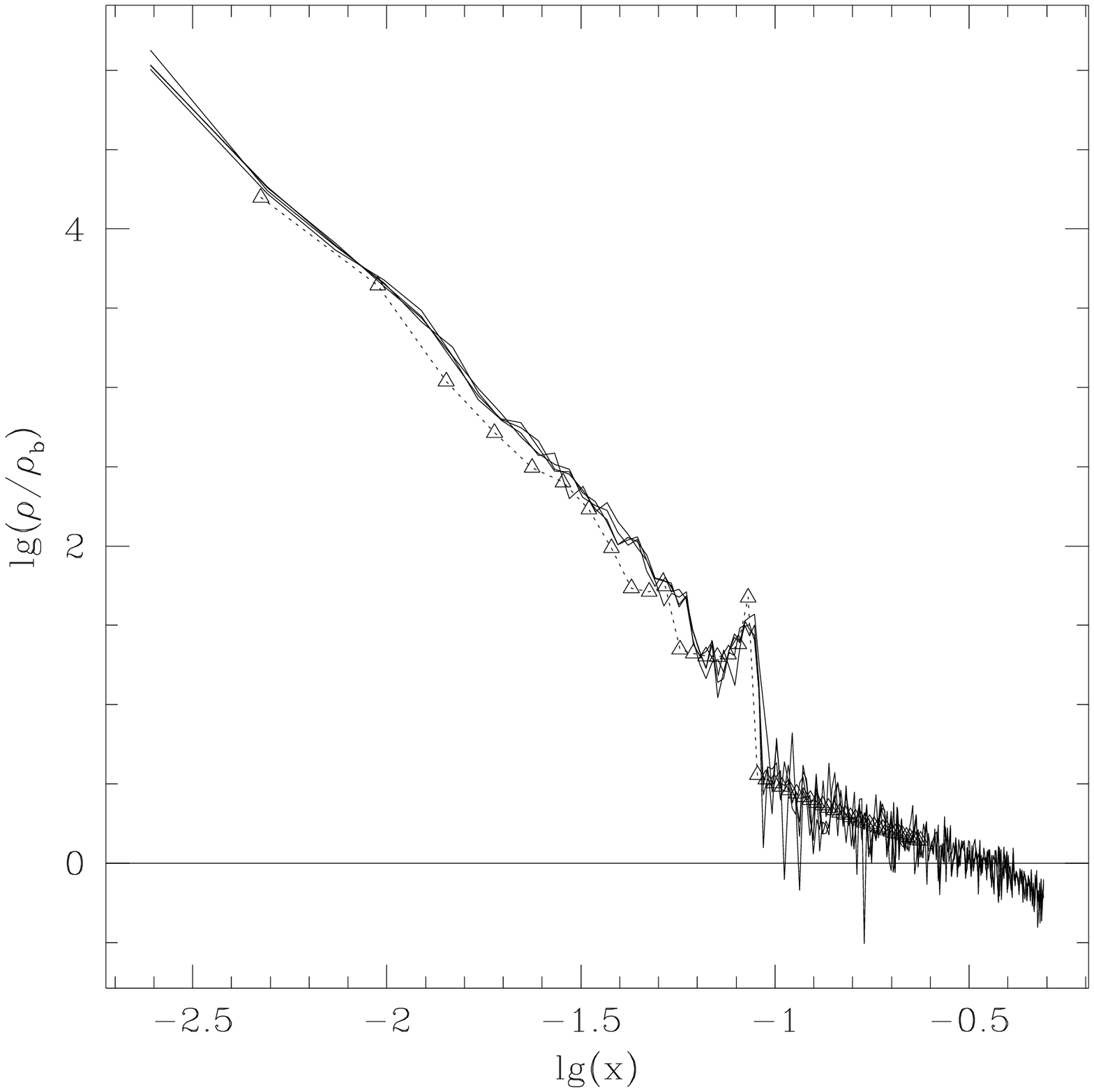]{Radial density profile after spherical infall.
The decimal logarithm of the density contrast $\delta = \rho/\rho_b$
is shown as a function of the logarithm of the comoving radius $x$,
for simulations of a sharp
peak accreting from a medium of uniform density $\rho_b$ for
an expansion factor $a_f/a_i=131$ in a critical universe.
The results of simulations realized with a single $64^3$ grid and with
a coarser $32^3$ grid and one subgrid of equivalent resolution to the
$64^3$ grid are overlaid. The results are insensitive to the details
of the treatment of particle flows through the subgrid boundary.
The open triangles and the dotted line that connects them are from
a table published by Bertschinger (1985), rescaled according to his
prescriptions.
The slope, the normalization, and the locations
of the first few caustics are in good agreement.
\label{f:tf6d-rho}}

\figcaption[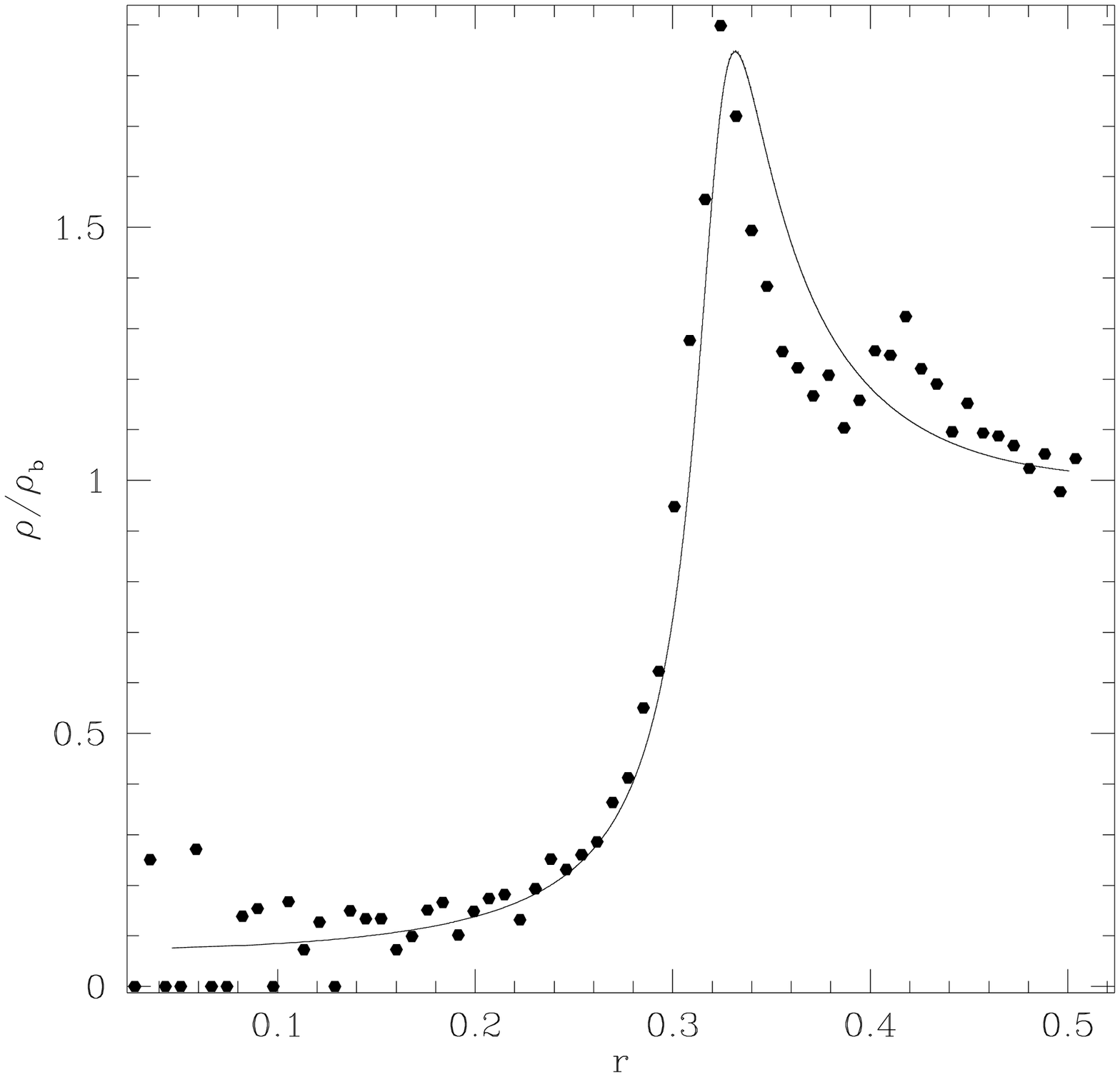]{Result of evolving a spherical hole with a
compensating ridge, as in Peebles (1987) and Peebles et al. (1989).
The density contrast (ratio of the local value to the mean value for
the cosmological model) is shown as a function of radial distance from
the center of the hole.
The curve is the analytic solution.
Points show the mean density in equally spaced radial bins, obtained
by running our code with a total of $64^3$ particles and $64^3$ grid
cells.
\label{f:ts6aa}}

\figcaption[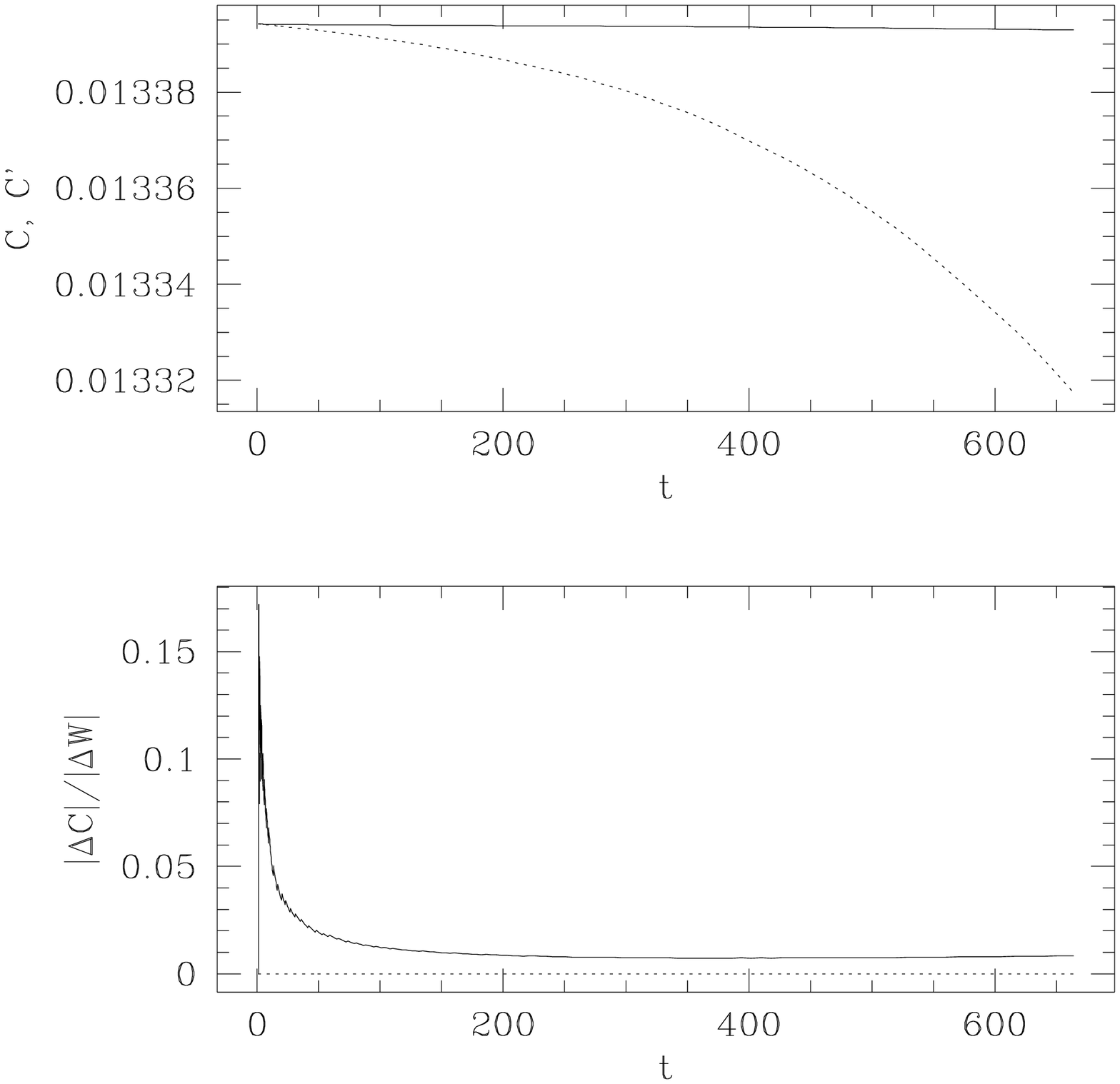]{Energy conservation in the expanding void
test. The top panel shows the evolution as a function of time ($t=3/2
(a/a_i)^{3/2}$) of the conserved quantities $C$
(equation~\protect\ref{q:C}, solid curve)
and $C'$ (equation~\protect\ref{q:Cprime}, dotted curve).
The bottom panel compares the change in $C$, respectively $C'$, to
that in $W$, respectively $aW$.
The initial transient is imputable to the round-off one expects when
both $|\Delta C|$ and $|\Delta W|$ are very small.
\label{f:ts6aa-econs}}
\end{document}